\documentclass[a4paper,fleqn,usenatbib]{mnras}

\usepackage{newtxtext,newtxmath}
\usepackage[T1]{fontenc}
\usepackage{ae,aecompl}

\usepackage{graphicx}	% Including figure files
\usepackage{amsmath}	% Advanced maths commands
\title[SN 1181 and Pa 30 from a Type {\rm I}ax Supernova]{The Path from the Chinese and Japanese Observations of Supernova 1181 AD, to a Type Iax Supernova, to the Merger of CO and ONe White Dwarfs}

\author[B. E. Schaefer]{
Bradley E. Schaefer$^{1}$\thanks{E-mail: schaefer@lsu.edu},
\\
$^{1}$Department of Physics and Astronomy, Louisiana State University, Baton Rouge, Louisiana, 70820, USA\\
}

\pubyear{2021}

\begin{document}
\label{firstpage}
\pagerange{\pageref{firstpage}--\pageref{lastpage}}
\maketitle

% Abstract of the paper
\begin{abstract}

In 1181 AD, Chinese and Japanese observers reported an unmoving bright `Guest Star' in the constellation {\it Chuanshe}, visible for 185 days.  In 2013, D. Patchick discovered what turned out to be a unique nebula surrounding a unique star, with the structure named `Pa 30', while subsequent workers made connections to mergers of white dwarfs, to the supernova subclass of low-luminosity Type {\rm I}ax, and to the 1181 transient.  Here, I provide a wide range of new observational evidence:  First, detailed analysis of the original Chinese and Japanese reports places the `Guest Star' of 1181 into a small region with the only interesting source being Pa 30.  Second, the ancient records confidently place the peak magnitude as 0.0$>$$V_{\rm peak}$$>$$-$1.4, and hence peak absolute magnitude $-$14.5$>$$M_{{\rm V,peak}}$$>$$-$16.0 mag.  Third, the Pa 30 central star is fading from $B$=14.9 in 1889, to $B$=16.20 in 1950, to $B$=16.58 in 2022.  Fourth, recent light curves show typical variability with full-amplitude of 0.24 mag on time-scales of one day and longer, critically with no coherent modulations for periods from 0.00046--10 days to strict limits.  Fifth, the spectral energy distribution from the far-infrared to the ultraviolet is a nearly-perfect power-law with $F_{\nu}\propto\nu^{0.99\pm0.07}$, observed luminosity 128$\pm$24 L$_{\odot}$, and absolute magnitude $M_{\rm V}$=$+$1.07.  I collect my new evidences with literature results to make a confident case to connect the East-Asian observations to a supernova, then to Pa 30, then to a low-luminosity Type {\rm I}ax SN, then to the only possible explosion mechanism as a merger between CO and ONe white dwarfs.
 
\end{abstract}

% Select between one and six entries from the list of approved keywords.
% Don't make up new ones.

%%%%%%%%%%%%%%%%%%%%%%%%%%%%%%%%%%%%%%%%%%%%%%%%%%

%%%%%%%%%%%%%%%%% BODY OF PAPER %%%%%%%%%%%%%%%%%%

\section{SUPERNOVA 1181}

The `ancient' observations of galactic supernovae (SNe) have proved to be valuable.  With the olden measures, the peak magnitudes, light curves, and even colours are central for physical models, and for making good connections between SN classes and supernova remnant (SNR) morphology.  Perhaps the most important reason for this is that the ancient observations provide a confident and accurate age for the modern models of the SNRs and pulsars, with these being critical for many physics issues.  Only five observed SNe have been confidently identified in the historical records; SN 1006 in Lupus, SN 1054 in Taurus (the Crab SN), SN 1181 in Cassiopeia, SN 1572 in Cassiopeia (Tycho's SN), and SN 1604 in Ophiucus (Kepler's SN){\footnote{All the other candidate SNe have uncertain identifications, to the point of not being useable for astrophysics.  For example, SN 185 is likely not a supernova at all, but is a good report of a famous comet at the same time and sky location (Schaefer 1995), while the supernova that created Cas A has never been confidently found in the historical records (SG2002)}.  Of these five confident SNe, four have well-confirmed SNRs.  The exception is SN 1181, which until recently had no confidently identified remnant.

SN 1181 (Stephenson \& Green 2002; SG2002) was discovered on 1181 August 6 in southern China, and then independently discovered over the next five days in Japan and in northern China.  The discovery record explicitly states that the `Guest Star' was visible for 185 days, and the last stated observation was on 1182 February 6.  The position was reported to be near the fifth star in the Chinese constellation of {\it Chuanshe}, which is in the north of the modern Cassiopeia, with all the Chinese and Japanese reports giving the same position and no indication of motion.  This long duration and the lack of motion prove that the 1181 event is what we would call a supernova.  

An identification for the modern counterpart of SN 1181 is what would provide the payoff for astrophysics.  The search for the counterpart has mainly been radio surveys looking for bright circular remnants and pulsars.  A number of radio remnants have been identified in the target area, however, all-but-one are certainly too old to associate with an SNR that is now 842 years old (SG2002).  But one SNR in the area was then reasonable, so Stephenson (1971) first suggested that the SNR 3C58 (G130.7+3.1) is the modern counterpart.  Later, many detailed physical studies have indicated that the 3C58 remnant is too old for SN 1181.  Fesen et al. (2008) measured the radial velocities of 450 optical emission knots to derive an age of $\sim$3000 years, while they also collect eight measures of the pulsar spin-down age, the proper motion expansion age, plus the ages from various models of the pulsar and its pulsar-wind-nebula, with a range from 2400--7000 years, a median of 3600 years, and no possibility that the age is 842 years.  So we were back to no recognized SNR.

Then in 2013, amateur astronomer D. Patchick, in hunting for planetary nebula candidates from WISE data, discovered what turned out to be a unique and extreme circular nebulosity surrounding a unique and extreme central star (Kronberger et al. 2014).  This nebula is catalogued as `Pa 30', being the 30th nebula discovered by Patchick, mostly in the {\it WISE} infrared sky survey.  Gvaramadze et al. (2019) made the recognition of the extreme and unique nature of the central star, plus made the connection to white dwarf mergers.  Oskinova et al. (2020) first recognized the unique and extreme nature of the central star, far past that of a simple-but-rare case of a very blue central star of a nebula, plus made the connection that the Pa 30 system is a supernova remnant from a Type {\rm I}ax event resulting from a merger of an ONe and a CO white dwarf.  Ritter et al. (2021) recognized that Pa 30 must have come from SN 1181, and that this opens the exciting program of making a close examination of an SNR with a known provenance.

The central star shines around 16th magnitude, in an uncrowded field at the centre of the nebula.  It had been previously catalogued as IRAS 00500+6713, with no recognition of its peculiarities until Patchick's discovery.  The star is highly luminous as seen with {\it XMM-Newton} in the X-rays, with {\it Galex} in the ultraviolet, and with {\it WISE} in the far infrared.   Optical spectroscopy shows prominent high-ionization emission lines (dominated by O {\rm VI}) with large widths, corresponding to wind velocities of 16,000 km s$^{-1}$ (Gvaramadze et al. 2019), with structure in the lines changing significantly in under 10 minutes, indicating a clumpy wind (Garnavich et al. 2020).  The reported surface temperature is 200,000 K, while the luminosity is $10^{4.5}$ L$_{\odot}$ (Gvaramadze et al. 2019).  The star is completely free of hydrogen and helium, while its surface composition is mostly carbon and oxygen. with substantial fractions of neon, magnesium, silicon, and sulphur (Gvaramazde et al. 2019; Oskinova et al. 2020).

The nebula is a circular, smooth, edge-brightened shell with an outer radius of 100$\pm$10 arc-secs.  The nebula is shining brightly in the far infrared, in X-rays, and in [O {\rm III}].  Optical spectroscopy gives an expansion velocity of $\approx$1100 km s$^{-1}$ and a kinematic age of 990$^{+280}_{-220}$ years (Ritter et al. 2021).  A detailed physics calculation of the wind from the central star ramming into the inner portion of the ejected shell gives an age of 350--1100 years (Oskinova et al. 2020).   X-ray spectroscopy shows that the nebula has no hydrogen or helium, contains 72 per cent carbon and all the remainder is oxygen, neon, and magnesium (Oskinova et al. 2020).  This hot gas is suffused with $\sim$0.008 M$_{\odot}$ of dust at a temperature of 60 K (Lykou et al. 2022).  

The presence of primarily carbon and oxygen, plus the neon and magnesium in large fractions, in both the central star and the nebula (Oskinova et al. 2020) forces the conclusion that the central star is some sort of a remnant from a catastrophic merger or explosion involving an ONe white dwarf (WD) and/or a CO WD.  And this catastrophic event must have been close to one millennium ago, and appeared in the old Chinese constellation of {\it Chuanshe}.  With this, the connection to SN 1181 is easy and likely.  The supernova event in 1181 AD cannot have been either a core collapse SN (CC-SN, like Type {\rm II}) or a thermonuclear explosion of a CO WD (Type {\rm I}a SN) because they cannot leave behind a remnant like that observed.  However, SN 1181 could have been a member of an uncommon subclass of SN called a `SN {\rm I}ax', that can leave behind a remnant as observed (Ritter et al. 2021).  

SN {\rm I}ax events `are the largest class of peculiar white dwarf (thermonuclear ) supernovae', characterized as being similar in spectra to Type {\rm I}a events, yet with lower ejecta velocities and lower luminosities (Jha 2017).  Many scenarios have been speculated, and we cannot say that we have any confident understanding of the system details.  Most of these models have in common a precursor involving a WD in a close binary, and some result in a remnant WD that is driving a fierce stellar wind.  This is sounding similar to the observed central star of Pa 30.  With this, the central star becomes the only closely observable remnant of a SN {\rm I}ax, and would become the touchstone for working out the nature of these events.

%SN 2012Z was a {\rm I}ax event in a nearby galaxy with deep imaging before and after its eruption, revealing a luminous blue source that is presumably the remnant star that survived the explosion of the white dwarf.  Four years after its explosion, SN  2008ha revealed a luminous red remnant left behind.  The spectral similarity to Type {\rm I}a SNe points to {\rm I}ax SNe as also being thermonuclear explosions of white dwarfs.  

\section{THE ANCIENT CHINESE AND JAPANESE DATA}

\begin{table*}
	\centering
	\caption{SN 1181 Positional Information from China and Japan}
	\begin{tabular}{ll}
		\hline
		Source & Positional description \\
		\hline
		
{\it Wenxian Tongkao}, {\it Songshi}	&		`appeared in {\it Kui} lunar lodge'	\\
{\it Wenxian Tongkao}	&		`invading {\it Chuanshe}', `beside the stars of {\it Chuanshe}', `guarded the star of {\it Chuanshe}', `in the guest house' 	\\
{\it Songshi} &	`invading the stars of {\it Chuanshe}'		\\
{\it Meigetsuki} diary	&		`guarding {\it Chuanshe}'	\\
SG2002	&		`a few days after discovery, the star was further said to guard ({\it shou}) the {\it fifth} star of {\it Chuanshe}'	\\
{\it Jinshi}	&		`seen at {\it Huagai}'	\\
{\it Meigetsuki} diary	&		`seen at the north near {\it Wangliang}'	\\
{\it Gyokuyo} diary	&		`beside {\it Ziwei}'	\\

		\hline
	\end{tabular}	
\end{table*}

\begin{figure}
	\includegraphics[width=1.0\columnwidth]{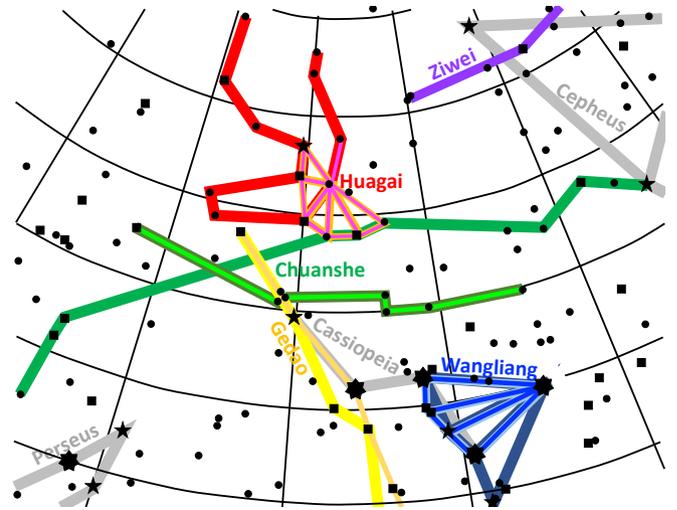}
    \caption{The Chinese constellations.  The constellations of {\it Chuanshe}, {\it Huagai}, {\it Wangliang}, {\it Ziwei}, and {\it Gedao} are shown with variations in their stars, as described in the text.  The modern constellations of Cassiopeia, Perseus, and Cepheus are depicted with thick gray lines, placed here to orient the readers.  The coordinates are for the year 1181 AD, north is up, east is to the left.  The declination circles run from $+$75$\degr$ for the small arc at the top, to $+$50$\degr$ across the bottom, at 5$\degr$ intervals.  The right ascension lines run from 22 hours on the right to 3 hours on the left, at one hour intervals.  Stars with $V$ magnitudes from 2.0--3.0 are depicted with large seven-pointed stars, 3.0--4.0 mag with five-pointed stars, 4.0--5.0 mag with fairly small squares, and 5.0--6.0 mag with small circles.  One point of this figure is to show the constellations that limit the position of SN 1181 (see Figs. 2--4).  Another point of this figure is to illustrate that included-stars in each of the constellations can have major uncertainties.}  
\end{figure}

The case centres on the old observations from China and Japan.  For this, we must realize that these old reports were not made by astronomers with a modern scientific mindset, but rather were made by high-level Confucian bureaucrats with a political and astrological mindset.  It actually matters for the astrological/political interpretation by the Southern Song astronomers that the guest star appeared at the gates of the celestial Imperial Palace (constellation {\it Ziwei}) at the time when an ambassador from the rival/enemy Jin empire arrived at the capital, whereupon the ambassador stayed in the Imperial guest house (constellation {\it Chuanshe}) until his departure from the Southern Song capital at the time when the supernova faded to invisibility.  As with historical reports from all times and regions, we must realize that the observations from 1181 might have rare errors{\footnote{For Chinese reports of old SNe, the most famous error is the statement that the Crab Supernova appeared just to the {\it south-east} of the star we now call $\zeta$ Tauri, whereas the Crab SNR is {\it north-east} of $\zeta$ Tauri.}, and that some aspects are poorly known and changed over time.  Despite these ordinary historiographic concerns, large scale studies have shown that the Chinese Imperial astronomical reports are almost always reliable in detail in their statements{\footnote{Many such studies are reported throughout SG2002, CS1977, and in other research papers of Stephenson.  These provide part of the strong reasons for knowing that the details in the Chinese records are of high reliability, and for taking the excellent work of Stephenson as being authoritative.}.

The best study of the Chinese records of SN 1181 is certainly the long and detailed chapter in SG2002, with this being authoritative and definitive.  The most recent discussion of these same Chinese records appears in Ritter et al. (2021), which unfortunately adopted most of its discussions and results from the thoroughly discredited source of Hoffmann, Vogt, \& Protte (2020), which has a high rate of bad errors of many kinds (Neuh\"{a}user \& Neuh\"{a}user 2021).  With this situation and with the new information on Pa 30 and supernovae, it is worthwhile re-examining the original Chinese and Japanese reports.

Fig. 1 shows the positions of the relevant Chinese constellations.  Each constellation is depicted with the stars from three authoritative sources, SG2002, Clark \& Stephenson (1977, CS1977), and Sun \& Kistemaker (1997, SK1997).  Each of the constellations has minor or major variations in the included stars, with these variations representative of the real uncertainties in the knowledge of these figures.

{\it Chuanshe}:  This constellation represents the Imperial Guest House just outside the celestial Imperial palace.  The old texts tell us that the constellation is nine stars running between {\it Huagai} and the Milky Way, but this leaves many possible paths, all involving faint stars.  SG2002 and CS1977 agree on the path, involving only seven stars, as shown by the broad band of dark green that has the narrower band of light green.  The path of SK1997 has nine stars and runs offset and roughly-parallel, as shown by the third broad line with a dark shade of green.  A critical and unresolved issue will be to identify the fifth star from the west in the constellation.  

{\it Wangliang}:  This constellation represents a famous charioteer, who is driving a team of four horses towards the east, with the four reins radiating out of the chariot.  The basic picture is well known, but the exact four stars for the four horses is open for discussion.  The SK1997 depiction is shown with broad bands of dark blue, the CS1977 depiction is shown with the narrow blue bands, while the SG2002 depiction is shown with the mid-sized light-blue bands.  

{\it Huagai}:  The canopy of the emperor, shading him while on the throne, has substantial uncertainty in the included stars.  The depiction in SG2002 (the broad orange lines) and CS1977 (the narrow magenta lines largely overlapping) show a parasol or umbrella-like figure.  Alternatively, SK1997 depicts the constellation (the broad red lines) as a throne with a chair-back and an overhanging canopy.  {\it Huagai} appears upside-down on this chart, but appears right-side-up when high in the skies above the pole as viewed from middle-latitudes.  

{\it Ziwei}: The Wall of the Imperial Palace is a large constellation, consisting of two arcs of stars surrounding the north polar regions that represent the palace grounds.  The only depiction of the relevant end is from SK1997 (thick purple line).  This constellation is one of the more important in the Chinese skies, separating out the Imperial Palace.  The Wall of the Imperial Palace has two gates (openings) situated on opposite sides of the pole, with one of the gates containing the Imperial Throne, and just outside that is the Imperial Guest House where ambassadors reside.

{\it Gedao}: This constellation depicts `a stepped road through mountainous territory' (SK1977, the broad tan line) or a `Hanging Gallery' (CS1977, the broad yellow line).  This constellation is not mentioned in conjunction with the SN of 1181, but it is a well-established group of bright stars that passes closely over the position of 3C58.  If the Guest Star of 1181 was at the position of 3C58, then the expectation is that the many ancient reports would have identified it as `invading' {\it Gedao} rather than {\it Chuanshe}, or at least have mentioned {\it Gedao}, although this is not a strong argument.

\subsection{The Sky Position of SN 1181}

The foremost question is the observed position on the sky of SN 1181.  For this, SG2002 constrains the position to be close to the east-west line of stars defining {\it Chuanshe} and inside the range in right ascension corresponding to the lunar lodge {\it Kui}.  In J2000 coordinates, this corresponds to a region roughly from 00:47 to 01:54 right ascension and from $+$63$\degr$ to $+$69$\degr$.  SG2002 favored the eastern side of this region, as based on one particular counting of the stars in {\it Chuanshe}.

The SN 1181 positional information from China and Japan is summarized in Table 1, with the translated texts from SG2002.  The sources are various surviving dynastic histories and several Japanese diaries.  The longest account is in the {\it Wenxian Tongkao}, which contains the astrological interpretation of the guest star appearing near the gate of the Imperial Palace as likening to the Jin ambassador coming to the Imperial Palace over the same time interval.  The {\it Wenxian Tongkao} is an encyclopedia likely based on the dynastic annals of the southern Sung, compiled around the year 1280, likely derived from the original daily records of the Song court at their capital of Linan (modern name Hangzhou).  We also have a brief report in the Song dynastic history {\it Songshi}, compiled around 1343, and clearly written from the same source material as {\it Wenxian Tongkao}.  In 1125 AD, Jurchen nomads conquered northern China from the Song dynasty, making their own Jin dynasty, which had its own dynastic annals, the {\it Jinshu}, compiled around 1343, with an independent set of observations.  From Japan, a history of the Kamakura Shogunate is called the {\it Azuma Kagami}, and this contains another set of independent observations.  The Japanese diaries are named {\it Meigetsuki}, {\it Kikki}, {\it Gyokuyo}, and {\it Hyakurensho}.  Table 1 does not include a variety of quotes for positional information that turns out to be not helpful (e.g., that the guest star `was seen at the north pole', which is to say that it was circumpolar for Japan).

\begin{figure}
	\includegraphics[width=1.0\columnwidth]{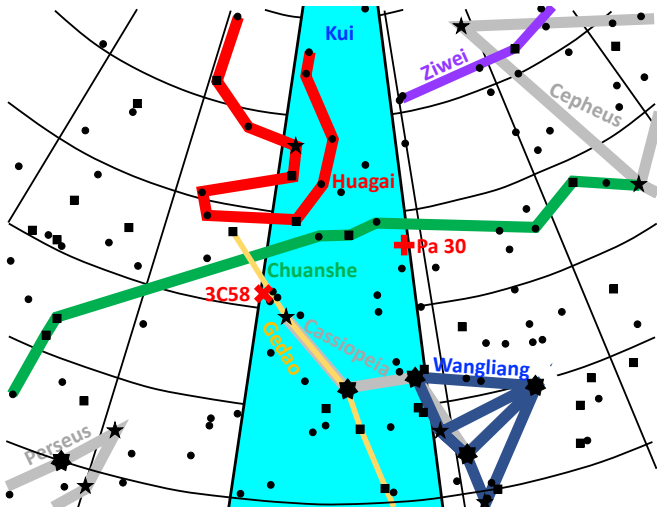}
    \caption{SN 1181 appeared inside the lunar lodge {\it Kui}.  The lunar lodge {\it Kui} represents an `orange-slice' region stretching from pole-to-pole, bounded by the right ascensions of the determinative stars, $\zeta$ And and $\beta$ Ari.  In the area around the modern Cassiopeia, the region of {\it Kui} is indicated by the cyan-shaded region.  The old candidate 3C58 is marked with a $\times$ symbol on the eastern edge, while the new candidate Pa 30 is marked by a $+$ symbol on the western edge.  The details of this star chart are the same as in Fig. 1, except that only the constellation outlines from SK1997 are shown.}  
\end{figure}

An important piece of positional information is that SN 1181 appeared inside the lunar lodge {\it Kui}, which is the orange-slice-shaped region of the sky bounded by the right ascensions of two `determinative stars'.  The 28 Chinese lunar lodges divide up the sky into slices, similarly as does the western zodiac.  The 15th lunar lodge of {\it Kui} has its edges defined by the right ascension of the stars $\zeta$ And (at right ascension 00:05 in the coordinates of 1181) and $\beta$ Ari (Sheratan, at right ascension 01:10 in 1181) for the determinative stars of SG2002 and SK1997.  Alternatively, CS1977 as well as Xu, Pankenier, \& Jiang (2000) give determinative stars of $\zeta$ And (1181 right ascension 00:14) and $\beta$ Ari.  The acceptable region inside the lunar lodge is represented as a cyan shaded area in Fig. 2.  This constraint has 3C58 at its far eastern side and Pa 30 at its far western side.

\begin{figure}
	\includegraphics[width=1.0\columnwidth]{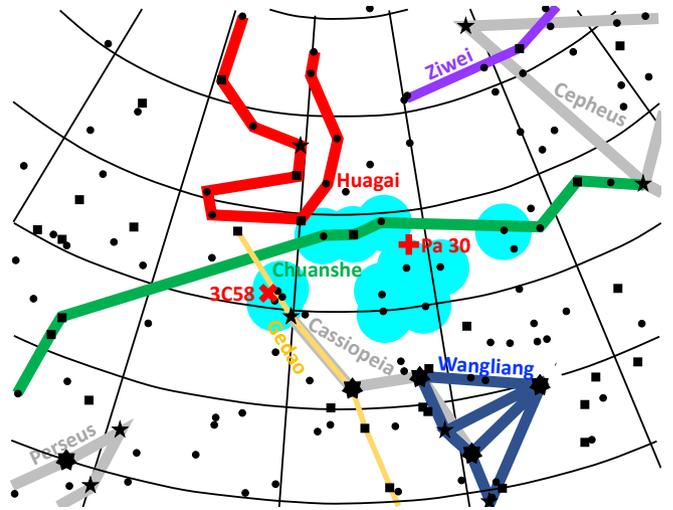}
    \caption{SN 1181 is close to the fifth star of {\it Chuanshe}.  {\it Chuanshe} was reported to be a line of nine stars extending east-west between {\it Huagai} and the Milky Way (which is roughly along the basic `{\rm W}' asterism of Cassiopeia).  The specific stars included in the Chinese constellation of {\it Chuanshe} are not known, and the fifth star from the west could be any one of eleven faint stars.  The indicated position for SN 1181 must be within something like 1.5$\degr$ of one of these 11 stars, as shown by the cyan-coloured circles centred on the stars.  The details of this figure are the same as for Fig. 2.}  
\end{figure}

Another strong positional constraint is that the guest star was `invading' and `guarding' the stars of {\it Chuanshe}, which is to say that the SN was within about 1 degree of one of the specific stars in {\it Chuanshe}.  Further, SG2002 says that the SN was guarding the {\it fifth} star of {\it Chuanshe}{\footnote{SG2002 does not provide a translation of the relevant text, apparently in the {\it Wenxian Tongkao}, nor can I find it in CS1977 or Xu et al. 2000.}, counting from the western end.  We do know that {\it Chuanshe} stretches in a row between the Milky Way (stretching along the modern `{\rm W}' asterism of Cassiopeia) and the Chinese constellation of {\it Huagai}, which likely has its southern edge near $\omega$ Cas.  The stars are nine in number and stretch in a long east-west line on the north edge of the modern Cassiopeia.  All of these stars are faint (below V=5.0), and there is no agreement as to which stars are in the set of nine.  Fig. 1 draws two reasonable sets of stars, one with seven stars from SG2002 and CS1977, and one with nine stars from the authoritative study of SK1997.  A problem with both depictions is that they incorporate stars that are faint, mostly from 5.1 to 5.9 mag, at which point the selection of stars to include must have been vague for the old observers and difficult to identify with any confidence for the modern scholars.  This uncertainty is compounded due to the lack of a reliable count of the star numbering to recognize which is the {\it fifth} star.  With just the seven stars depicted in SG2002, the fifth star is 53 Cas ($V$=5.61) close to 3C58.  But if the 8th and 9th stars of {\it Chuanshe} are added on the west side, then the fifth star would be 32 Cas (constant at $V$=5.57).  Alternatively, the fifth star of the Sun \& Kistemaker depiction is 36 Cas ($\psi$ Cas, at $V$=4.71).  Given the uncertainties in the constituent stars, other possible fifth stars include 43 Cas ($V$=5.57), 31 Cas ($V$=5.32), HR 177 ($V$=5.82), HR 233 ($V$=5.36), HR 273 ($V$=5.96), HR 342 ($V$=5.55), HR 567 ($V$=5.28), and HR 9104 ($V$=5.69).  So SN 1181 is within a degree or so of one of these 11 stars, and we have no useful evidence to confidently pick out which of these stars.  Fig. 2 depicts these constraints as multiple overlapping cyan-coloured circles with radius 1.5$\degr$.

\begin{figure}
	\includegraphics[width=1.0\columnwidth]{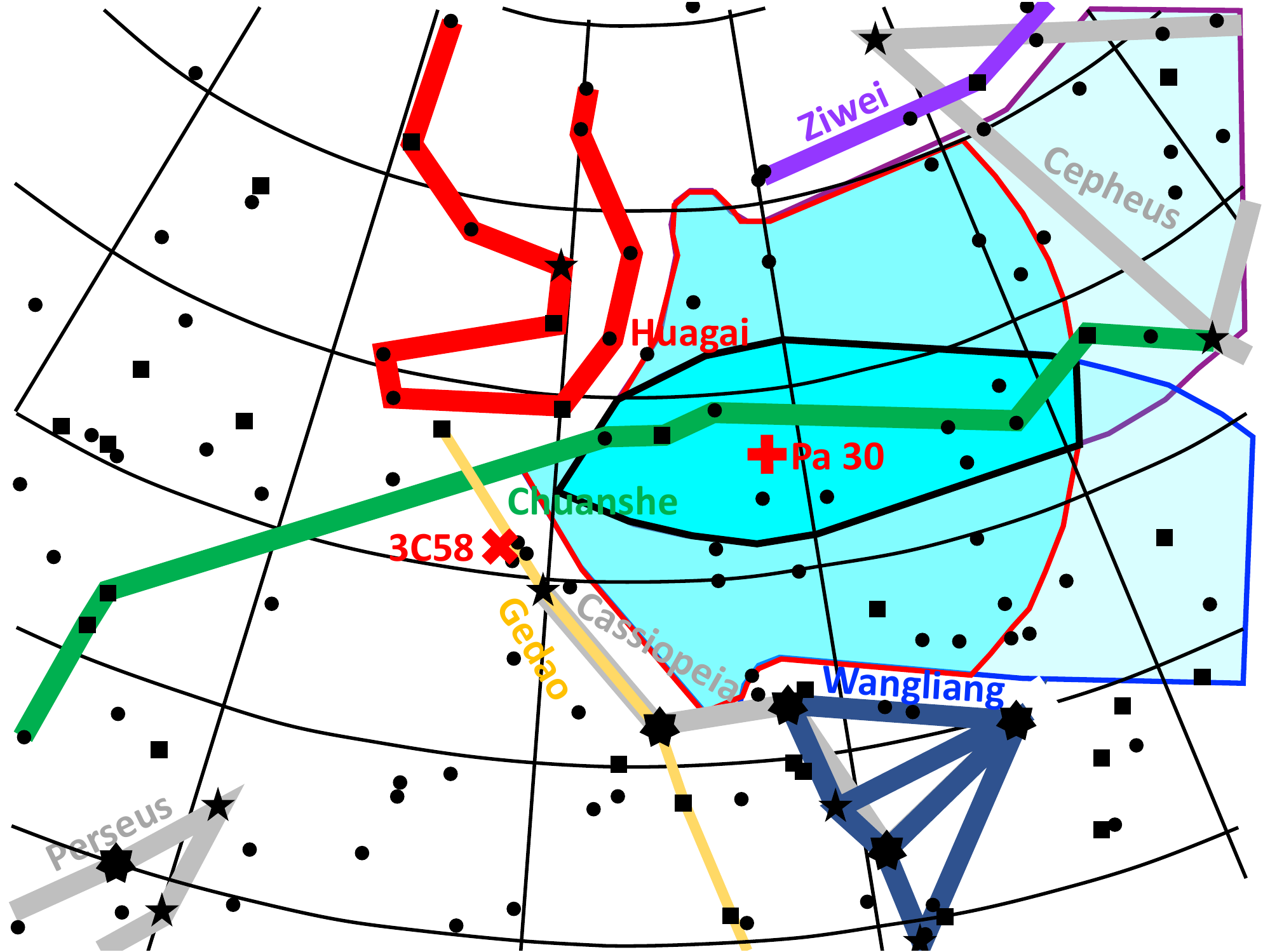}
    \caption{SN 1181 is near to {\it Wangliang}, {\it Huagai}, and {\it Ziwei}.  The meaning of the words `near {\it Wangliang}', `at {\it Huagai}', and `beside {\it Ziwei}' point to SN 1181 being near but outside the constellation.  Unfortunately, the proximity is fairly vague.  With the supernova being described as near both {\it Ziwei} and {\it Wanglian}, separated by 14$\degr$, the upper limit must be $>$7$\degr$.  For the constraints in this figure, I have adopted 10$\degr$ as the upper limit for proximity, but this limit has substantial uncertainty.  For {\it Huagai}, this limit must be applied to the westernmost star in the SG2002 depiction, i.e., the allowable position must be within 10$\degr$ of 31  Cas.  A further constraint arises because the supernova was outside the constellation, which I take to be a distance of greater than 1$\degr$ from its stars.  Further, I constrain the reports of nearness to mean that the supernova position was not inside or past some other constellation.  The resultant allowed positions for being near to {\it Wangliang} is shown as a faint cyan-shaded region with a blue border.  The allowed region for the proximity to {\it Huagai} is a faint cyan-shaded region with a red border.  The region `beside {\it Ziwei}' is displayed as a cyan-shaded region with a purple border.  The regions that satisfy only one constraint are outlined in faint cyan, while the regions with two overlapping constraints are in a darker cyan.  All three constraints overlap in one central area, with a bright cyan colour and a thick black border.  That is, all three proximity constraints are only satisfied by the lens-shaped region with Pa 30 near its centre.  The details of this figure are the same as for Figs. 2 and 3.}  
\end{figure}

\begin{figure}
	\includegraphics[width=1.0\columnwidth]{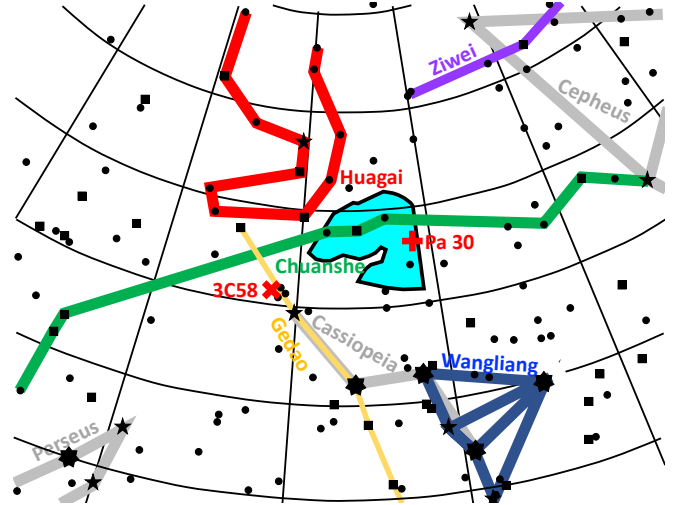}
    \caption{SN 1181 position from all constraints.  The three sets of constraints (the deep cyan regions in each of Figs. 2, 3, and 4) form a joint constraint on the SN 1181 position, as shown by the cyan region with black edges in this figure.  The details of this figure are the same as for the earlier figures.  The final shape is convoluted because of the circular constraints from Fig. 3 (the nearness to candidate fifth stars of {\it Chuanshe}).  The final error region has an area of roughly 15 square-degrees.  The position of Pa 30 is near one of the edges, but this is fine as the Chinese astronomers would have reported such a guest star as being inside {\it Kui} whether it is near the edge or near the centre.  The main takeaway from this figure is that Pa 30 is definitely inside the small final positional region for SN 1181.}  
\end{figure}

The other useful positional evidence is that the supernova was seen near each of the three Chinese constellations {\it Huagai}, {\it Wangliang}, and {\it Ziwei}.  In all three cases, the Chinese word implies that the guest star was near to the constellation, but not too near.  For {\it Huagai}, SG2002 state `With regard to the position of the guest star, the term {\it yu} (`at') is rather vague and merely denotes general proximity to {\it Huagai}.'  With {\it Ziwei} and {\it Wangliang} being 14$\degr$ apart at closest, the proximity apparently extends out past 7$\degr$.  For a lower limit, the supernova position cannot be nearer than something like 1$\degr$ to the stars of a constellation, or the report would have reported the guest star to `guard' or `invade' the constellation.  I have adopted a criterion for the proximity that the supernova position must be from 1$\degr$ to 10$\degr$ of any star for the proposed constellations.  These proximity alerts are useful because these limit the plausible position of the SN.  In Fig. 4, I have constructed three regions, representing the allowed supernova positions for the proximity to each of the constellations.  The triple overlap is a fairly large lens-shaped area with Pa 30 near its centre.

We have three sets of positional constraints taken from the old Chinese and Japanese data.  These are depicted as cyan-shaded regions in Figures 2, 3, and 4.  The position of SN 1181 must be inside the intersection of these three regions (see Fig. 5).

One quick realization is that SN 1181 cannot be from 3C58 because it has proximity with neither {\it Wangliang} nor {\it Ziwei}.  Indeed, 3C58 is separated from {\it Ziwei} by two independent constellations, so there is no chance that the Japanese diarist would have called the supernova as being `beside {\it Ziwei}, with its nearest star 12.1$\degr$ distant.  This positional proof from the East-Asian astronomers provides another rejection of 3C58, with this being more primary and fundamental than the strong argument that 3C58 is much too old.

An important point for later is that no known SNR, other than Pa 30, is inside this error box (Green 2019).  For example, Tycho's SNR is nearby, but it violates the criterion of being inside {\it Kui}, it violates the loose constraint of proximity to {\it Ziwei}, and it violates the known date of 1572 from Tycho.  Two old SNRs appear just to the south of the final region in Fig. 5 (Green 2019), G126.2+1.6 (70 arc-minutes in size) and G127.1+0.5 (45 arc-minutes in size).  Both SNRs appear close enough to HR 342 (a candidate fifth star of {\it Chuanshe}), but they are more than 11$\degr$ from {\it Ziwei}.  Given the vagueness of the reported position as `beside {\it Ziwei}', an advocate could expand the allowed region to include these two SNRs.  However, for any reasonable distance and average velocity, both remnants are greatly older than tens of thousands of years.  With this, there are no SNRs that can be associated with SN 1181 -- other than Pa 30.

Pa 30 is inside the allowed region.  This strongly makes the connection from the supernova event of 1181 and the modern SN remnant.

%That is to say, the ancient Chinese and Japanese positional data are consistent only for a small area of the sky, and this includes the position of Pa 30.  

\subsection{The Peak Magnitude of SN 1181}

The second big question for the ancient data concerns the peak magnitude.  With the modern measures of distance and extinction, the observed peak magnitude can be converted into an absolute magnitude at peak, $M_{{\rm V,peak}}$.  This is then diagnostic of the SN type and the energetics of the event.

SG2002 give the best analysis of the observed peak magnitude, $V_{\rm peak}$, but this can be updated with some additional information on supernova light curves.  That astronomers in three nations discovered the SN within five days of each other demonstrates that the initial discovery was likely made at a time of a fast rising light curve.  Further, this demonstrates that the magnitude at the time of discovery must have been first magnitude or brighter, with a peak magnitude then of $V_{\rm peak}$=0 or perhaps much brighter.  The {\it Wenxian Tongkao} record the star as `large', again pointing to it being bright.  The {\it Azuma Kagami} says that the guest star `had rays', which means that it is sufficiently bright for the ordinary spiked visual defects in all human eyes to rise above the threshold for visibility.  The limiting magnitude for visible rays is ill-defined and varies from observer-to-observer, but SG2002 conclude that the SN must have been brighter than $V$=0.  However, the ancient records do not record any daytime visibility or comparisons with the Moon, Venus, or Jupiter, so there must be a weak upper limit that $V_{\rm peak}$$>$$-$5 or so.  An ambiguous constraint arises because the {\it Azuma Kagami} compares the star to the planet Saturn.  At the time of discovery, Saturn was only briefly visible in the dawn sky at magnitude $+$0.4, while months after peak when Saturn is near opposition, its magnitude is 0.0.  But the comparison to Saturn cannot return any confident brightness estimate because some translations have the comparison being made to the {\it colour} of Saturn, while it is possible or likely that any brightness comparison was made at a time long after the peak of the supernova.  So we are left with a range for $V_{\rm peak}$ from roughly 0.0 mag to roughly $-$5 mag.

With the knowledge that we are dealing with a supernova, we can take modern light curve information to transform the duration of visibility (185 days) into a peak magnitude.  The critical line of reasoning is that both the Song and Jin astronomers closely followed the SN, and it disappeared from notice on day 185 or day 156 after the original discovery, respectively.  Neither the lunar phase nor the seasonal gap affect these dates of disappearances, so that means that these dates are when the SN faded to some limit of recognition.  SG2002 gives good evidence that this limit cannot be fainter than $+$5.5 mag, while inattentive observers could easily have given up at even $+$3.2 mag{\footnote{This is the magnitude at which the human eye flips between photopic and scotopic vision for stars in a dark night sky (Schaefer 1993).  This is the threshold where the supernova would no longer be visible by looking directly at the position with direct vision, rather the visibility must be with averted vision.}.  So the supernova light curve was at 3.2--5.5 mag at a time of 185 days after discovery.  The three independent discoveries were all made in a five night interval, and this points to the first day being on the fast rising portion of the pre-maximum light curve, or roughly 10 days before peak.  With the peak around day 10 and its disappearance on day 185, the SN was lost close to 175 days after peak.  Now we can take modern SN light curves and look to see how many magnitudes below peak it is around day 175.  The fastest observed cases are for ordinary Type {\rm I}a events, for which the decline will be by near 4.8 mags.  In this extreme case, the peak would be 4.8 mag brighter than 3.2--5.5 mag, or in the range of $+$0.7 to $-$1.6 mag.  But this number is just to set the stage, because we know that Pa 30 cannot be a Type {\rm I}a SN.  Type {\rm I}ax SNe fade more slowly, with declines over 175 days ranging from 3.1 to 4.6 mag (McCully et al. 2014b).  For one extreme, the range of $V_{\rm peak}$ is then from $+$0.1 to $-$1.4 mag, while the other extreme has the peak from $+$2.4 to $+$0.9 mag.  That is, for any sort of  Type {\rm I}ax SN light curve, $V_{\rm peak}$ must be from $+$2.4 to $-$1.4 mag so as to disappear roughly 185 days after the initial discovery.  And we still have the constraint 0.0$>$$V_{\rm peak}$$>$$-$5 from the Chinese and Japanese descriptions.  The joint constraint is that SN 1181 must have peaked between 0.0 and $-$1.4 mag.

\subsection{Distance, Proper Motion, and Peak Absolute Magnitude}

The distance to Pa 30, $D$, has been previously quoted as 3070$^{+340}_{-280}$ pc (Gvaramadze et al. 2019), 3100 pc (Oskinova et al. 2020), and 2300$\pm$140 pc (Ritter et al. 2021).  I can update this now with the {\it Gaia} DR3 parallax of 0.406$\pm$0.026 milli-arc-secs{\footnote{https://gea.esac.esa.int/archive/}.  The fractional error is sufficiently small such that the traditional and simple calculation is accurate, thus giving a distance of 2460$\pm$160 pc.  In principle, the parallaxes should be calculated with a Bayesian equation and priors appropriate for the galactic population (Bailer-Jones 2015).    We have no useful priors other than that from the galactic latitude ($+$4.63$\degr$) and the appropriate exponential scale height for a young stellar population of 150 pc feeding into the `exponentially decreasing space density' (EDSD) prior as recommended by the {\it Gaia} Team (Luri et al. 2018).  My calculation gives the best {\it Gaia} distance of 2410$^{+200}_{-130}$ pc.

{\it Gaia} DR3 gives the total proper motion of 2.766$\pm$0.038 milli-arc-secs (mas) per year.  The only way to get an estimate of the pre-SN proper motion is to look at nearby stars of similar parallax, in the hope that the SN progenitor was co-moving with this population.   I have constructed a diagram plotting the observed proper motions in right ascension and declination for 85 stars within 360 arc-secs of Pa 30 and with parallax from 0.30--0.50 mas, with these forming a diffuse pattern with an RMS scatter of around 2 mas per year.  The central star of Pa 30 differs from this proper motion vector by 1.3 mas per year, so there is no significant evidence for any substantial kick on the stellar remnant.  With the questionable assumption that the SN progenitor is co-moving, the SN kick on Pa 30 can only estimated as 1.3$\pm$2.0 mas per year, with this translating at the {\it Gaia} distance to be a transverse velocity of 15$\pm$23 km s$^{-1}$.   

The extinction, $E(B-V)$, has been reported as 0.8 mag from fitting the spectral energy distribution (Gvaramadze et al. 2019), 0.84$\pm$0.04 mag from optical spectroscopy (Oskinova et al. 2020), and 0.70 mag from three methods (Ritter et al. 2021).  I confirm that these are reasonable values, because Pa 30 is 198 pc above the galactic plane at the {\it Gaia} distance, which places it above something like 87 per cent of the total galactic extinction along that line of sight, which Schlafly \& Finkbeiner (2011) measure to be 1.09 mag, for an approximate estimate of 0.95 mag.  I will adopt $E(B-V)$ of 0.84$\pm$0.04 mag.

For the two extreme cases of peak magnitude, I calculate absolute magnitudes of $-$14.55$\pm$0.18 and $-$15.95$\pm$0.18.  This can be represented by saying that the best estimate value of $M_{{\rm V,peak}}$ is $-$15.25, while the extreme allowed range is close to $-$14.5 to $-$16.0.

This $M_{{\rm V,peak}}$ is greatly less luminous than the Type {\rm I}a SNe, and it is in the bottom portion of the luminosity distribution for Type {\rm I}ax SNe.  Jha (2017) lists $M_{{\rm V,peak}}$ values observed for 51 Type {\rm I}ax SNe.  Only five SNe are less luminous than my best estimate for SN 1181.  That is, SN 1181 is in the bottom 10 per cent of Type {\rm I}ax SNe.

\section{LIGHT CURVE FROM THE LAST CENTURY}

The elemental abundances of the central star and the kinematic age of the expanding nebula demonstrate that it started out as some sort of a supernova $\sim$842 years ago.  Such a remnant must have started out hot and bright.  As the remnant loses energy from radiation and its wind, the remnant can only fade.  So I predict that the central star has been fading significantly over the last century.  If we can look back one century ago, the remnant star should have been substantially brighter, as 100 years is a substantial fraction of its current age of 842 years.  This fading after a millennium of time has also been predicted for some specific models of SN {\rm I}ax models (Shen \& Schwab 2017).  

\subsection{The Archival Sky Photographs at Harvard}

The only source of photometry from more than 25 years ago is archived sky photographs (called `plates') that happen to go deep enough to record the target star.  The majority of all archival plates worldwide are in the collection now at the Harvard College Observatory (HCO) in Cambridge Mass, with roughly 500,000 plates covering the entire sky, north and south.  The HCO plates cover the interval of 1890 to 1954, plus coverage from the late 1960's up to 1989, and are largely the only source of plates before the 1920's.  The current $B$ magnitude of the central star is close to 16th mag, and so it can be recorded only on the best and deepest of the century-old plates.  In particular, for plates with limiting magnitudes of 16th mag from before 1950, the only source is the HCO archives.

%\footnote{The only exception to this is that the plate archives at Sonneberg Observatory have some small coverage that goes deep back into the 1930s.}.  

The magnitude of the central star is measured from the image diameter of the target star in comparison with nearby stars of known magnitudes.  The traditional method, used from the 1890s to the present, to compare star diameters is for an experienced observer to visually examine star pairs when seen with a loupe on a light table.  For the usual good situations, an experienced observer will measure target magnitudes to a one-sigma uncertainty of $\pm$0.10 mag.  I have a long and deep experience with measuring target magnitudes, as well as experiments and theory for the measurement process (e.g., Schaefer 1981; 2016; 2020; Schaefer \& Patterson 1983).  My experiments show that my by-eye measures have the same measurement uncertainty and zero offset from the measures made by modern photometry programs based on scanned plates.  Further, my visual measures are confidently better than the best of the scanned photometry for a variety of critical situations.  For the central star of Pa 30, the critical situation is that the scan-based computer photometry often fails to recognize targets when near to the plate limit, thus returning no measure for plates where the by-eye measurements return a confident and accurate magnitude.

The DASCH  (Digital Access to a Sky Century @ Harvard) program (Grindlay et al. 2012) has already digitized many of the deepest plates and these are available on-line\footnote{http://dasch.rc.fas.harvard.edu/search.php}.  For the Harvard plates, the image of the Pa 30 central star is isolated and far above the plate limit.  To illustrate this, Fig. 6 shows extracted regions for two plates, with these scans produced by the DASCH system.  The important point for this pair of plates is that the central star has changed substantially in brightness from 1924 to 1950, and this is easily seen by comparing the star images to nearby comparison stars.

\begin{figure}
	\includegraphics[width=1.0\columnwidth]{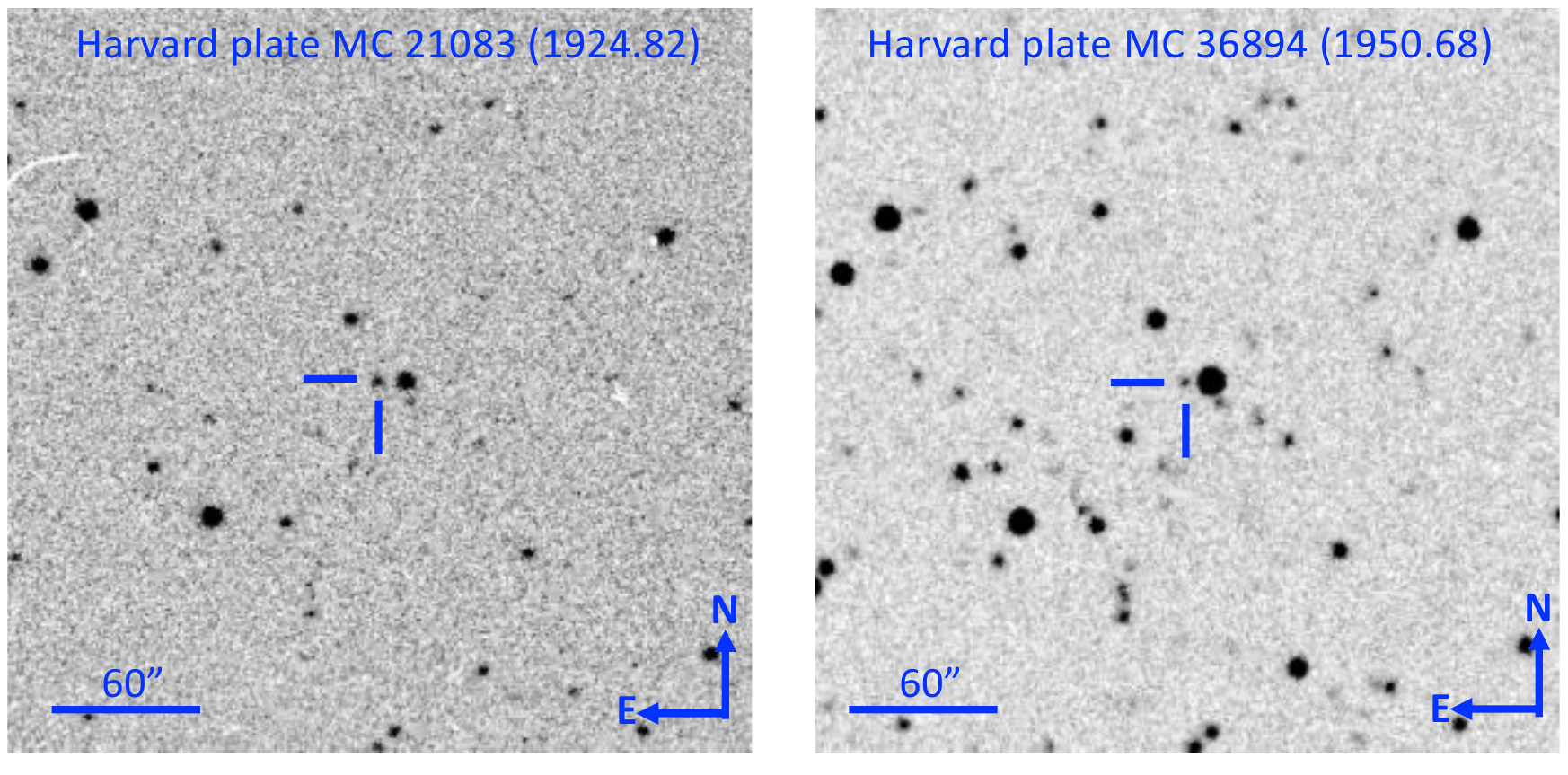}
    \caption{Harvard plates from 1924 and 1950.  The archival plate collection at HCO contains many old and deep plates, with 43 plates from 1889--1950 having good detections of the central star of Pa 30.  This figure shows closeups (roughly 5' by 5' square) of two of the plates, with north up and east to the left, and with the central star indicated by a pair of tick marks.  Part of the reason to show these two plates is to illustrate that the images of the central star are of good quality, from which we can get reliable magnitudes.  Part of the reason to show these plates is because the large drop in magnitude is readily seen.  For example, compare the central star with the constant star just south of the bright star to the west, in 1924 the central star was much brighter than the comparison star, while in 1950 the comparison star was somewhat fainter than the comparison star.}  
\end{figure}

The DASCH program has already measured $B$-magnitudes of the central star for 22 plates in the years 1924--1950.  Critically, I have chosen the DASCH comparison stars to be calibrated with the APASS $B$ magnitudes.  For the simple case of an isolated star, the DASCH magnitudes are reliable, and the average reported photometric uncertainty for the central star is $\pm$0.15 mag.  Further, I have used the DASCH scans of these 22 plates to get by-eye magnitude estimates by visual comparison of the central star versus nearby comparison stars.  Further, I have visited the Harvard archives and visually examined and measured many additional plates that have not been scanned.  This allows a substantial extension of the light curve back to the year 1889.  This also allows recognition of flaws missed in any scans, for example, the plate MC 12053 has a small circular dent at the position of the central star that is proven to be an artifact by side illumination of the emulsion.  For each plates, I have 2--5 independent measures of the magnitude, and I have averaged these together so as to beat down the measurement errors.  In the end, I have extended the DASCH light curve (22 plates 1924--1950) to 46 plates from 1889--1950, all with somewhat better magnitudes due to multiple independent measures.

In the end, I have 43 good measures (plus 3 useful limits) of the $B$-band brightness from 1889 to 1950.  The one-sigma uncertainty on these measures is 0.10--0.15 mag.  These magnitudes are tabulated in Table 2 and plotted in Fig. 7.  For the Harvard plates, the last column of Table 2 gives the Harvard plate number in parentheses.

\begin{table}
	\centering
	\caption{Light curve 1889--2022}
	\begin{tabular}{llll}
		\hline
		Julian Date & Year  &  $B$ (mag)  &  Source \\
		\hline
		
2411323.6	&	1889.88	&	15.10	&	HCO (I 144)	\\
2413113.7	&	1894.78	&	14.77	&	HCO (A1047)	\\
2413422.7	&	1895.63	&	$>$15.1	&	HCO (A1579)	\\
2413431.7	&	1895.65	&	15.40	&	HCO (A1614)	\\
2421165.7	&	1916.83	&	$>$15.5	&	HCO (MC 11454)	\\
2421186.6	&	1916.89	&	$>$15.3	&	HCO (MC 11691)	\\
2421546.6	&	1917.87	&	15.08	&	HCO (MF 1308)	\\
2423297.8	&	1922.66	&	16.09	&	HCO (MC 18951)	\\
2424081.6	&	1924.81	&	15.45	&	HCO (MC 21064)	\\
2424084.7	&	1924.82	&	15.48	&	HCO (MC 21083)	\\
2424388.8	&	1925.65	&	15.70	&	HCO (MD 21572)	\\
2424388.8	&	1925.65	&	15.74	&	HCO (MC 21572)	\\
2424471.5	&	1925.88	&	15.60	&	HCO (MC 21702)	\\
2424472.6	&	1925.88	&	15.76	&	HCO (MC 21706)	\\
2424770.8	&	1926.70	&	15.62	&	HCO (MC 22082)	\\
2424790.7	&	1926.75	&	16.06	&	HCO (MC 22117)	\\
2424824.6	&	1926.84	&	15.76	&	HCO (MC 22191)	\\
2424872.5	&	1926.98	&	15.80	&	HCO (MC 22289)	\\
2425149.8	&	1927.73	&	15.83	&	HCO (MC 22712)	\\
2425179.7	&	1927.82	&	15.53	&	HCO (MC 22827)	\\
2425208.6	&	1927.90	&	16.10	&	HCO (MC 22932)	\\
2425258.5	&	1928.03	&	15.95	&	HCO (MC 23061)	\\
2425528.7	&	1928.77	&	16.03	&	HCO (MC 23707)	\\
2425557.6	&	1928.85	&	15.89	&	HCO (MC 23788)	\\
2425620.5	&	1929.02	&	15.88	&	HCO (MC 23974)	\\
2425977.5	&	1930.00	&	15.95	&	HCO (MC 24681)	\\
2426251.7	&	1930.75	&	15.85	&	HCO (MC 25145)	\\
2426592.8	&	1931.69	&	15.91	&	HCO (MC 25609)	\\
2426653.7	&	1931.85	&	16.08	&	HCO (MC 25738)	\\
2427357.7	&	1933.78	&	15.60	&	HCO (I 52391)	\\
2427361.7	&	1933.79	&	15.69	&	HCO (I 52412)	\\
2427361.8	&	1933.79	&	15.55	&	HCO (I 52414)	\\
2427417.6	&	1933.94	&	15.56	&	HCO (I 52538)	\\
2428102.6	&	1935.82	&	15.90	&	HCO (RL 637)	\\
2428461.6	&	1936.80	&	15.86	&	HCO (MC 28527)	\\
2428461.6	&	1936.80	&	15.97	&	HCO (MD 28527)	\\
2430227.8	&	1941.64	&	16.11	&	HCO (IR 4998)	\\
2430250.5	&	1941.70	&	16.13	&	HCO (IR 5065)	\\
2430264.6	&	1941.74	&	15.98	&	HCO (IR 5133)	\\
2430264.6	&	1941.74	&	16.04	&	HCO (IR 5134)	\\
2430264.8	&	1941.74	&	15.84	&	HCO (IR 5136)	\\
2430676.5	&	1942.87	&	16.19	&	HCO (IR 6332)	\\
2430676.5	&	1942.87	&	16.04	&	HCO (IR 6333)	\\
2431033.6	&	1943.84	&	16.26	&	HCO (MC 33153)	\\
2432465.7	&	1947.76	&	16.19	&	HCO (IR 10262)	\\
2433531.8	&	1950.68	&	16.53	&	HCO (MC 36894)	\\
2435015.9	&	1954.75	&	15.84	&	POSS I	\\
2455855.9	&	2011.80	&	16.44	&	APASS	\\
2455946.7	&	2012.05	&	16.33	&	APASS	\\
2455949.7	&	2012.06	&	16.40	&	APASS	\\
2455968.6	&	2012.11	&	16.24	&	APASS	\\
2455983.6	&	2012.15	&	16.29	&	APASS	\\
2459817.6	&	2022.65	&	16.545	&	Vanmunster	\\
2459818.6	&	2022.65	&	16.670	&	Vanmunster	\\
2459819.6	&	2022.66	&	16.515	&	Vanmunster	\\

		\hline
	\end{tabular}	
\end{table}

\begin{figure}
	\includegraphics[width=1.01\columnwidth]{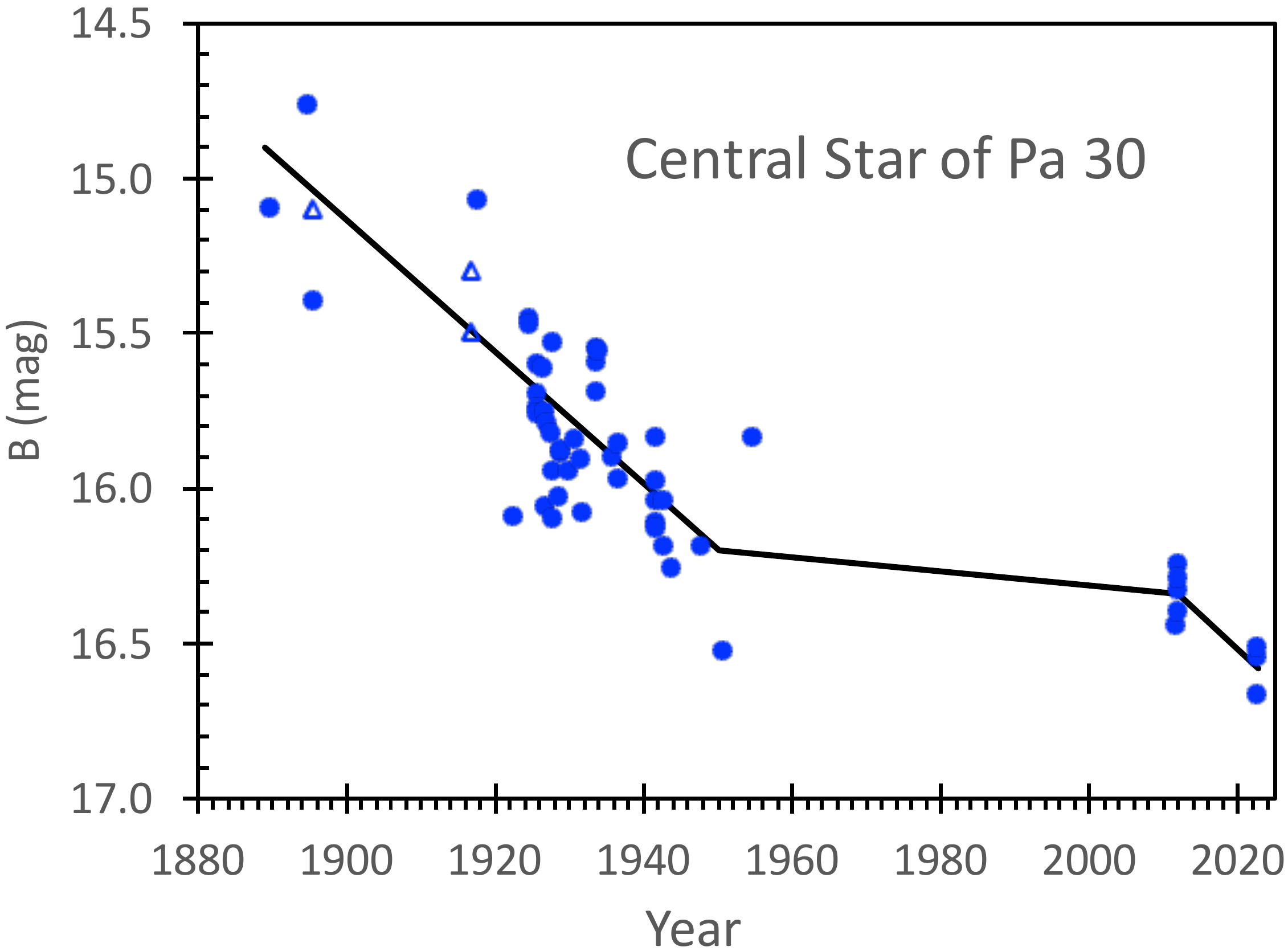}
    \caption{The central star is fading fast from 1889--2022.  This light curve is from Table 2, with typical error bars of 0.10--0.15 mag before 1960, and $\pm$0.03 for the last two decades.  All data sets have the observed scatter around the mean light curve being substantially larger than the measurement errors, so the scatter is dominated by the intrinsic night-to-night variability of order 0.2 mag.  The primary point of this figure is to test my prediction that a surviving stellar remnant of an 842 year old supernova might still be fast fading over the past century.  To test this prediction, it is vital that all the $B$ magnitudes be calibrated into an identical photometric system, and this renders useless the large amounts of post-1950 photometry.  The result of this test is that the stellar remnant is certainly fading at a fast rate, with a drop of 1.68 mag in the last 105 years.}  
\end{figure}

The light curve from the Harvard plates alone (from 1889 to 1950) shows a fast fading of the central star, from $B$=14.9 in 1889 to $B$=16.2 in 1950.  This is highly significant, and the change in brightness is readily apparent from simply looking at the plates (see Fig. 7).  So we already have our answer; the central star has been fast fading over the last century, at least by 1.3 mags from 1889--1950.

The Harvard magnitudes show a substantial scatter around any smooth curve.  For the broken line fit to the 1889--1950 light curve (see Fig. 7), the RMS of the deviations equal to 0.19 mag.  This is significantly larger than the real measurement error of 0.10--0.15 mag.  This is undoubtedly due to the intrinsic variability of the central star on time-scales of one-day and longer.  With the deviations arising from independent measurement error and intrinsic variability, the half-amplitude of the $B$-band intrinsic variability is roughly $\sqrt{0.19^2 - 0.15^2}$=0.12, for a full peak-to-peak amplitude of 0.23 mag.

An important point, usually overlooked by the inexperienced, is that we have to be careful about colour corrections so that all the photometry is placed on to the same colour-system with the same calibrations.  For the case of the Harvard plates, the spectral sensitivity is effectively identical to that of the Johnson $B$-magnitudes, as well it should because the Harvard photometry was the immediate predecessor that was used for defining the modern system.  For defining the magnitude system, a further critical issue is the zero-magnitude of the scale, as defined by the adopted comparison stars..  Both DASCH and I used the same APASS (AAVSO Photometric All-Sky Survey, Henden et al. 2009) calibration of the comparison star magnitudes\footnote{https://www.aavso.org/download-apass-data}, and these are in the Johnson $B$-system.  With this, both the DASCH magnitudes and my magnitudes are exactly in the modern $B$-magnitude system.

Lykou et al. (2022) also used the DASCH magnitudes for the central star.  They used plates from 1924--1950, and reported that the central star was fading at a rate of 2 mag per century.  They only  used DASCH photometry, so they did not include half of the useable plates.  Their result has the critical problem of using the `GSC2.3.2 Catalog' for the calibration of their comparison stars, with these magnitudes being in the `$Jpg$' system fairly close to $V$.  Their adopted comparison stars use magnitudes greatly different from the native $B$-system of the HCO plates, so there will necessarily be large and unknowable colour corrections that vary from plate-to-plate and year-to-year and that vary with the intrinsic changes of the spectral energy distribution and the line fluxes.  The result is that their light curve will display systematic offsets that shift over time, creating apparently-linear secular trends, with these looking similar to the fading reported by Lykou et al. (2022)\footnote{The reason why secular trends are introduced as artefacts is because the Harvard plates have systematic changes in limiting magnitudes, plate scales, and emulsion types over the decades, such that the comparison stars and their color-corrections to go from $B$ magnitudes to $Jpg$ magnitudes have a systematic shift over the years.  These effects in the DASCH light curves are easily seen for the Pa 30 central star by noting that the trend reported by Lykou et al. with the GSC calibration completely goes away when the comparably-bad ATLAS calibration is used, demonstrating that bad-calibrations can make apparently significant secular trends come and go.  The same error has been made by other researchers who simply downloaded DASCH light curves and selected bad comparison stars, with this leading to false secular trends and wrong science claims in various cases (e.g., see Schaefer et al. 2018).  The lesson is that only the APASS calibration can be used if good long-term photometric accuracy and stability are needed.}.

\subsection{Towards a Consistent Light Curve from 1889--2022}

The light curve from the Harvard plates already answers the question as to whether the central star is fading, but we really should also get the light curve from 1950--2022.  This turns out to be much harder than most people would expect.  The trouble is that most of the more modern data cannot be placed on to the same magnitude system as the Harvard magnitudes, or at least to a confident accuracy of better than 0.10 mag or so.  That is, most post-1950 data are in bands with substantially different spectral sensitivity (e.g., $V$, $g$, $R$, $R_C$, $zr$, unfiltered, ...) and the zero magnitude  is on a different scale (e.g., AB magnitudes, Vega magnitudes, ...) from that used for the Harvard plates.  For the purposes of creating a single consistent light curve from 1889--2022, all the magnitudes must be on a single specific colour system and calibration, and this must be that used for measuring the Harvard plates.  That is, to create a single homogenous light curve, the post-1950 magnitudes must have a native spectral sensitivity close to the Johnson $B$-system, and it must be calibrated with comparison stars from APASS or an equivalent measure.

In practice, it is not possible to correct other systems to produce a $B$-magnitude to the needed accuracy of better than 0.10 mag or so.  Further, it is impossible to know the colour correction, e.g., $B-V$, for a target with variable colours and strong variable spectral lines (Gvaramadze et al. 2019; Garnavich et al. 2020; Lykou et al. 2022).

Further, these problems are ubiquitous even for photometry in the various `blue' or nominally-`$B$' passbands.  For photometry with a system that is close to the spectral sensitivity of the Johnson $B$ system (say, with a Bessell $B$ filter, or with 103a-O emulsion), the problem of getting a consistent zero-magnitude is always present.  The trouble is that all the sources of comparison stars have offsets with respect to each other (and to alternate versions of the same catalog).  With long experience (e.g., Schaefer et al. 2011; 2018), I find typical offsets of 0.1 mag and up to 0.4 mag.  So for purposes of constructing a century-long light curve accurate to better than 0.10 mag, I can only use photometry whose native system is close to the Johnson $B$-band spectral sensitivity and that are calibrated with comparison stars from the APASS survey. 

\subsection{Other Archival Plate Sources}

The central star is easily visible on the blue plate from the first Palomar Sky Survey (POSS I), dating to 1954 September 30.  The emulsion is 103a-O, for which the native system is close to the Johnson $B$-system.  I measured the brightness of the central star as compared to nearby stars both by-eye and with the measures in the USNO-B catalog (Monet et al. 2003), getting closely similar values.   The comparison stars are from APASS, so this 1954 magnitude is a modern $B$ magnitude.  With this, $B$=15.84$\pm$0.10.  

The Maria Mitchell Observatory archival collection\footnote{https://www.mariamitchell.org/astronomical-plates-collection} has 8513 plates from 1913--1995, with typical limiting magnitude of 14--16.  The catalog reveals five plates that cover the target with sufficient exposure time to be hopeful.  On viewing scans of all five plates, the limiting magnitudes are all too bright to show the central star.

Oskinova et al. (2020) report one approximate $V$ magnitude from two archival plates from 1926 at the Hamburg Observatory.  The native system of these plates is the same as at Harvard (because everyone was using similar emulsions), so it is unclear how a $V$ magnitude was derived, and any such magnitude cannot be compared with any other data source with any useable accuracy.  More importantly, the point-spread-function (PSF) of the nearby bright star overlaps with the PSF of the central star, covering up over two-thirds of its PSF diameter, which is to say that not enough of the target star is showing to make a useable brightness estimate.  Further, the faint edge of the target star's PSF is only a marginal detection, of questionable significance.  In all, the Hamburg plates return nothing useful for this target.

The POSS II plate from 1994 was taken with {\rm III}aJ emulsion plus a GG395 filter, with this magnitude not convertible to the photometric system of the Harvard plates.  For plate sources that go deep, no useful plates are found for the Tautenburg Schmidt, the Kitt Peak Schmidt, and the Carnegie Observatories telescopes in California.

\subsection{CCD Magnitudes}

The average magnitudes for comparison stars are available from the APASS survey through its AAVSO webpage.  The individual magnitudes for the central star, as observed with APASS, are also publicly available.  These magnitudes are for the $B$, $V$, $g'$, and $r'$ passbands, with these all being useful for constructing the spectral energy distribution (Section 5).  The five $B$-band measures are listed in Table 2.  These are over several months in 2011/2012.  The five measures, on different nights, range over 0.20 mag, which is greatly larger than the photometric uncertainty.  I interpret this as being due to the known intrinsic night-to-night variability

At my request, T. Vanmunster observed the central star on three nights (2022 August 26--28) with a 0.40-m f/5.1 Newtonian telescope located in Landen Belgium, observing with a Bessell $B$ filter.   Vanmunster is one of the most experienced and capable observers of cataclysmic variables in the world.  The calibration of his photometry was made using APASS comparison stars, so his magnitudes are directly comparable to the APASS magnitudes of 2011/2012 and the Harvard magnitudes of 1889--1950.  Vanmunster took 27 images, so the scatter in these shows us the real measurement error bars.  The nightly averages are given in Table 2.  The uncertainty in the quoted nightly averages is 0.03 mag.  The three nightly averages differ from each other by much more than the measurement errors, so the variability is intrinsic to the central star.

Other sources of CCD magnitudes, all from after the year 2000, cannot be inter-compared or accurately placed on my long-term light curve due to the problems of inter-comparing observations with different spectral sensitivities and different calibrations of comparison stars for a target with a greatly unusual and fast-variable spectrum.  This includes all the photometric sources collected by Gvaramazde et al. (2019) and  Lykou et al. (2022).

\subsection{Light Curve From 1889--2022}

The final light curve from 1889--2022 is shown in Fig. 7.  The intrinsic variability of $\sim$0.2 mag is seen in all data sets.  For an answer to the original question, we see that the central star of Pa 30 is certainly fading and fading fast.  I have drawn in some lines fitted to the light curve, with this intended only to be a description of the light curve in hand.  From 1889 to 1950, the central star faded at the unprecedentedly huge rate of 2.1 mag per century.  The gap from 1950 to 2011 is filled with the modest decline rate of 0.2 mag per century.  From 2012 to 2022, the decline rate has increased to 2.2 mag per century.  Such high decline rates lasting for over a century are completely unprecedented for any known source.

A further result from this fading is that the supernova must have happened only some several centuries before 1889.  Taking the fade rate of 2.1 magnitudes per century from 1889 to 1950, the central star would be magnitude 0.0 in the year 1180 AD.  Such extrapolations can be made from rates over various time intervals inside the 1889--2022 range, and pushed to target $V$ magnitudes variously of $-$1.4, 0.0 or 3.2--5.5, with a wide range of derived ages.  All such linear extrapolations must have poor accuracy, because we already see a highly variable fade rate 1889--2022 and because theory models for remnant winds predict large variability (Shen \& Scwab 2017).  Nevertheless, the fade rate alone gives an age for the remnant that is one millennium, to order-of-magnitude accuracy.

With the further information that the real age is currently 842 years, the {\it average} fade rate from $V$=5.5 in the year 1182 down to $V$=14.9 in 1889 is 1.3 mag per century.  The fade rate was presumably fast in the first years after 1182, then it must have been relatively slow in later centuries, only to start a fast fade as observed in the years 1889--1950, followed by a slower-yet-variable fading until the current time.  Just this sort of situation for the fade rate has previously been predicted for low-luminosity Type {\rm I}ax WD remnants (Shen \& Schwab 2017).  In particular, their models (see their fig. 8) show a fast fading over the first few years after 1181, then a nearly flat light curve for many decades-to-centuries, with a sudden drop followed by a slowing in the fade rate, all with the approximately correct luminosity at the start and end.

\section{FAST PHOTOMETRIC VARIABILITY}

\subsection{Variability on Day-to-Month Time-scales}

\begin{figure}
	\includegraphics[width=0.95\columnwidth]{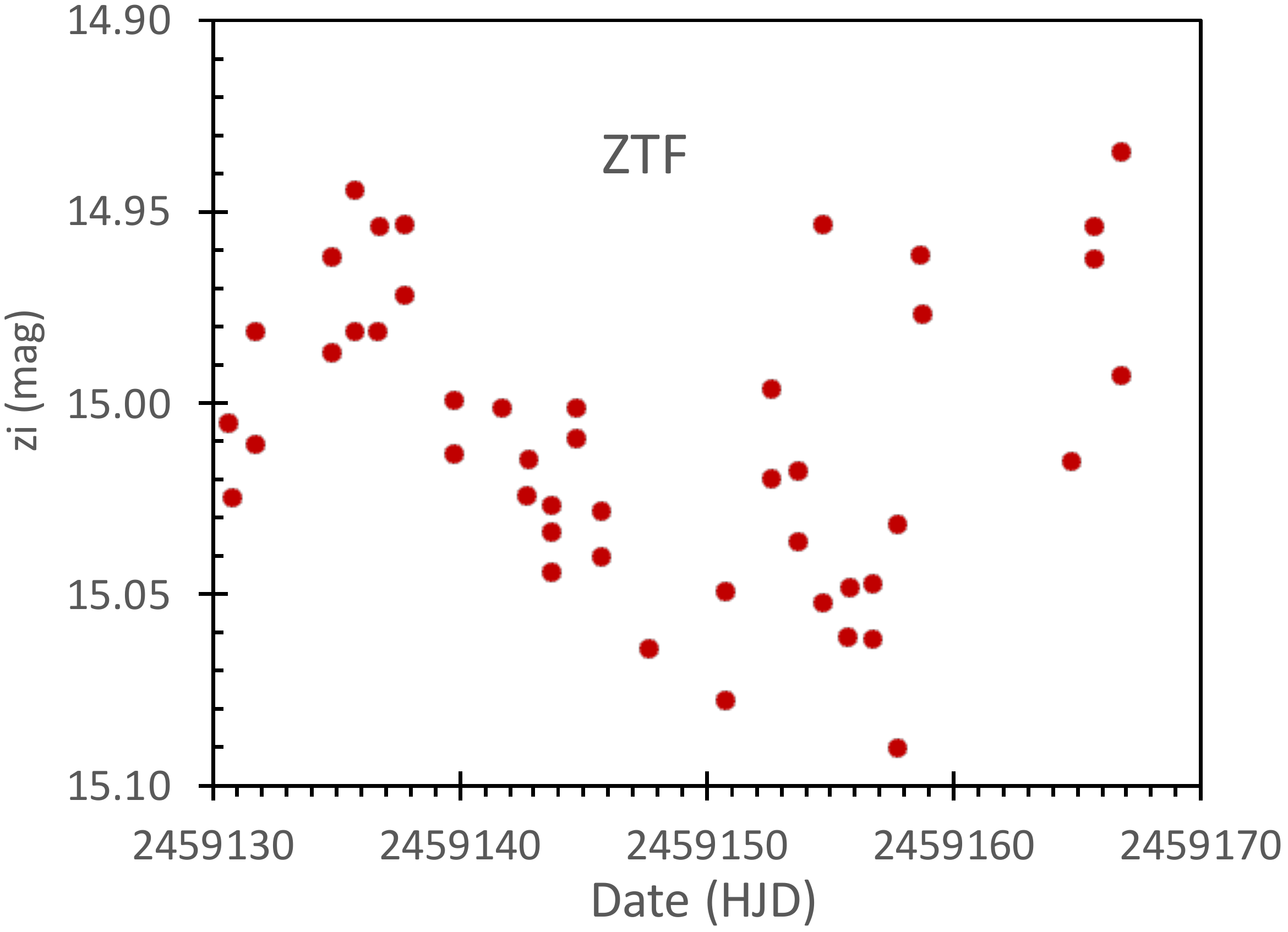}
    \caption{ZTF light curve for the central star.  This shows the central star's typical variability, with non-periodic variability on time-scales from one day to one month, looking chaotic.  The error bars on these deep-red magnitudes are $\pm$0.011 mag.}  
\end{figure}

The {\it Gaia} light curve has a substantial variance, with an estimated range of 0.08 mag, as based solely on the {\it Gaia} quoted uncertainty in the mean flux (Chornay et al. 2021).  For their science of seeking orbital variations in central stars of planetary nebulae, Chornay et al. {\it assumed} the variability to be periodic, and concluded that the central star is a `candidate binary'.  But the variability is not periodic (see below) so this claim for binarity is not applicable.

The Zwicky Transient Factory (ZTF) has been producing light curves on many nights for the entire sky north of $-$30$\degr$ declination for many years since 2018 (Bellm et al. 2019).  For the case of the central star of Pa 30, ZTF has 46 measures within 36 days in 2020 October/November\footnote{https://irsa.ipac.caltech.edu/cgi-bin/Gator/nph-scan?projshort=ZTF}, but only in one band, the {\it zi} band covering 7100--8700~\AA.  This light curve (see Fig. 8) shows two vague peaks separated by a month, with four flares lasting 1--2 days and with $\sim$0.05 mag amplitude.  The total amplitude of variability is 0.16 mag from the extremes, or 0.10 mag with smoothing.

In Section 3, I have three measures of the variability on time-scales from days to months:  On three successive nights, Vanmunster found the central star to vary by up to 0.15 mag.  Over a four month interval, on five nights, APASS found variability in the $B$-band of up to 0.20 mag.  From 1889--1950, the 46 Harvard plates show a scatter above the variance from the measurement errors that corresponds to a full-amplitude of 0.23 mag.  These measures just sample the variability, so the long-term amplitude will be somewhat larger.

The {\it TESS} light curve provides a long nearly-continuous light curve (see next Section and Fig. 8), and this provides a full-sampling of the underlying range of variations.  With smoothing to remove the Poisson variations, Fig. 9 shows variations from 120--150 counts/sec, which is a 23 per cent variability, a 0.24 mag full amplitude.

\subsection{{\it TESS} Photometry}

\begin{figure*}
	\includegraphics[width=2.1\columnwidth]{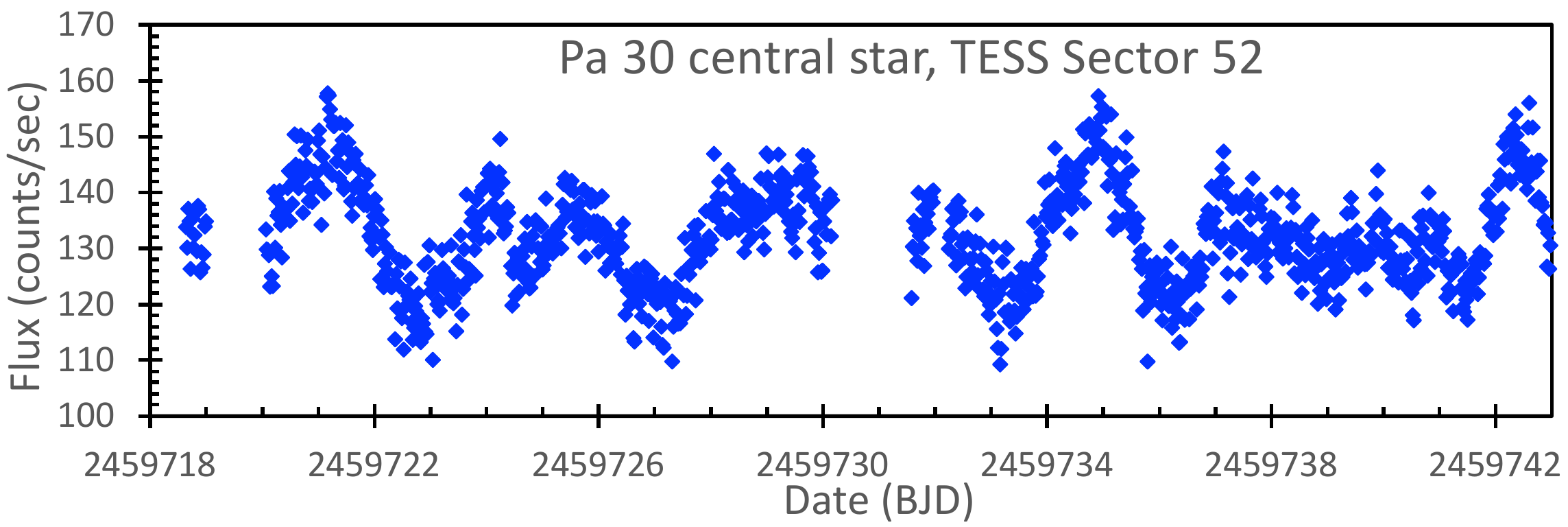}
    \caption{{\it TESS} light curve for the central star.  This background-subtracted light curve is for Sector 52 (2022 May 18 to June 13), with 1800 second time resolution.  The one-sigma error bars are $\pm$2.6 counts/sec, so the scatter around a smooth curve is just from ordinary Poisson variations.  The fastest significant variability has a time-scale of roughly 0.5 days, the typical rise-and-fall time-scale is near 3 days, and no long-term trend is apparent.  Critically, a Fourier transform of the {\it TESS} light curves shows no significant coherent periodicity for periods ranging from 40 seconds up to 10 days.}  
\end{figure*}

The {\it TESS} satellite is designed to return excellent and continuous time series photometry for most stars in the sky down to roughly 19th magnitude (Ricker et al. 2015).  The {\it TESS} light curve and Full Frame Images are publicly available on-line by the Mikulski Archive for Space Telescopes (MAST)\footnote{https://mast.stsci.edu/portal/Mashup/Clients/Mast/Portal.html}.  The images are unfiltered and cover roughly 6000-10,000~\AA.  The light curves have few gaps and have integration times from 20 seconds up to 1800 seconds, depending on the date and whether the specific star has been targeted for fast photometry.  The photometry is broken up into `Sectors', where each Sector covers a 24$\degr$$\times$90$\degr$ field of view for a duration of around 27 days (two orbits, with a small gap between).  For the case of Pa 30, {\it TESS} has already covered it during four Sectors (Sectors 18, 24, 25, and 52) from 2019 November to 2022 June.  Pa 30 was not targeted during the first three Sectors, so the light curve integration times were 1800 seconds, which is the cadence for sending down the Full Frame Images.  For Sector 52, Pa 30 was specifically targeted, so the spacecraft sent down Target Pixel Files for a small region around the target with a 20 second cadence.  So we have nearly a hundred days of continuous photometry (with two long gaps and five short gaps) with 20 second and 1800 second time resolution.  Such a data set is good for detecting periodic photometric variations with periods of under 10 days or so.

The {\it TESS} light curve shows irregular variability on all time-scales from half a day up to one week and longer (see Fig. 9).  For time-scales faster than 1800 s (available for Sector 52 only), the light curve is dominated by Poisson fluctuations and shows no significant variability.  Fig. 9 shows the Sector 52 light curve binned up to 1800 s time resolution.  For time-scales from 1800 s to 0.5 days, the light curve is still dominated by Poisson noise, with no significant intrinsic variations.  By eye, there appears no eclipses, no sinusoidal modulations, and no periodic regularity.  The central star of Pa 30 is 21.9 arc-seconds directly east of a star 4.6 mag brighter in the red, so with 21 arc-second pixels for TESS, the SPOC photometry aperture includes roughly 19 per cent of the neighboring starlight, and dominates the background.  The Poisson noise from all the flux is $\pm$2.4 e/s for the 1800 s binning in Fig. 9.  The extra light from the nearby star is a very stable fraction of the star's light, which is not variable, and produces no peaks in the Fourier transform.  The quoted limits on the amplitude of coherent modulations are made with this nearby starlight included.  In all, the extra light from the nearby star only makes a modest increase in the noise in the Fourier transforms, and there is no means for this extra light to hide any coherent periodicity of the central star of Pa 30.  

I have calculated Fourier transforms for the {\it TESS} light curves, seeking coherent periodic modulations with periods from 40 seconds to 10 days.  No significant periods were identified.  The limits on the amplitude of modulation can be identified by looking at the upper envelope of the noise peaks in the Fourier power spectra.  For orbital periods from 0.6 to 10 days (i.e., those for a sub-giant companion filling its Roche lobe), the full-amplitude must be less than 5.7 counts/sec, or 0.043 mag.  For orbital periods from 0.04 to 0.6 days (i.e., for any unevolved star near to filling its Roche lobe), the full-amplitude must be less than 0.0045 mag.  For periods from 40 seconds to 0.04 days (i.e., a rotation period for the white dwarf), the full-amplitude must be less than 0.014 mag.  These strict limits can be used to reject the likelihood of any nearby companion star.

\subsection{Can a Binary Companion Star be Hidden?}

Several of the models considered below have the SN progenitor with a consequential companion star, where the star serves as a donor for accretion, the second star for a merger, or as a means to strip a massive star of its outer layers.  Further, the progenitor might have had an inconsequential companion kept at a distance from the exploding star and of no import for its explosion.  Any companion star will survive largely unchanged through the SN event (Marietta et al. 2000), unless it was part of some in-spiral merger, so it should still be visible.  The existence and nature of any companion star is an observable property that can distinguish between models for the explosion mechanism.

Some models can have the companion star being a giant star or a sub-giant star.  However, any giant or sub-giant companion star would be revealed in the spectral energy distribution (SED), where the usual blackbody shape would be easily recognized sitting on top of the power law from the stellar wind (see Fig. 10).  No blackbody is seen, with limits down to near 0.1$\times$ the power-law flux.  The central star has an accurately known absolute magnitude of $+$1.07 mag, so the absolute magnitude of the companion star must be fainter than approximately $+$3.6 mag.  Giant stars always have absolute magnitudes more luminous than $+$1.4 mag, while sub-giants are brighter than $+$3.2 mag.  Thus, inconsequential-or-consequential binary companions of the giant or sub-giant classes are not in the Pa 30 system.  

The limit from the SED (M$_V$$>$$+$3.6 for any surviving companion) only allows for a companion star less massive than our Sun, for stars on the main sequence.  If such a companion star is distant from the exploding progenitor, then it must be inconsequential, and will not be helpful for distinguishing between models.  To distinguish models, we must test for a nearby companion that was at some time in contact with its Roche lobe.  Any M$_V$$>$$+$3.6 binary companion that provides material for Roche lobe overflow would have an orbital period between 0.04 and 0.6 days.  All close binaries have some level of photometric variability on the orbital period (e.g., from eclipses, irradiation effects, tidal effects, and ellipsoidal effects), while accreting binaries have additional mechanisms for photometric modulations on the orbital period (e.g., the strongly-asymmetric beaming pattern from a hotspot).  These orbital modulations must be present at some level, and {\it TESS} is the perfect tool for discovering the periodicity.  

The {\it TESS} light curve has no coherent periodicities to strict limits, so the conclusion is that the central star of Pa 30 is {\it not} a consequential binary.  This result will come to the forefront near the end of this paper, when the binarity issue decides between the only two acceptable models.  With this importance, it behooves us to look for loopholes in the no-companion result.  I can think of four generic ideas to hide a consequential companion star:

{\bf (1)} If the plane of the orbit is nearly perpendicular to the line of sight, the amplitude of modulation will be small.  The amplitude will scale approximately as $A_{90\degr} \sin[i]$, where $A_{90\degr}$ is some constant for scaling the variability, and $i$ is the inclination of the orbit (close to 0$\degr$ for the orbit's pole pointing close to Earth).  The scale factor for any binary in Pa 30 is not known.  But for cases of X-ray binaries and cataclysmic variables, the typical modulation on the orbital period is 0.1 mag for a 45$\degr$ inclination, so $A_{90\degr}$=0.14 mag.  In such a case, the central star of Pa 30 would need an orientation of the orbital plane pointing to within 1.8$\degr$ of Earth to keep the modulations under the 0.0045 mag limit.  The probability of such an accurate pointing towards Earth is unlikely at the 0.00049 probability level.  

{\bf (2)} Perhaps the periodic signal can be hidden by minimizing the periodic modulation so as to make the amplitude below the 0.0045 mag limit?  For example, a model might require that the accretion has stopped after the SN in 1181.  For this, there is an irreducible minimum $A_{90\degr}$ caused by the reflection effect of the hot WD making the inward facing hemisphere of the companion star shine with a brightness that will modulate on the orbital period.  For all consequential binaries, the mass ratio (the mass of the companion star divided by the mass of the WD) is approximately unity, for which the companion would cover roughly 4 per cent of the luminosity emitted by the primary.  This would lead to 0.04 mag variations for edge-on systems, for $A_{90\degr}$=0.04 mag.  Any such periodic modulation would be easily and surely picked up by {\it TESS}.  This minimal modulation can be hidden from Earth if we are looking nearly pole-on to the binary orbit.  To make the minimal modulation appear with less than 0.0045 mag full-amplitude, the orbit must point to within 6$\degr$ of Earth, with the probability of such a close coincidence being 0.005 (i.e., nearly a three-sigma rejection).  So the periodic signal from the companion star can be minimized only so far, and such is still unacceptably unlikely for explaining the {\it TESS} light curve.   

{\bf (3)} Perhaps the consequential companion star became unbounded in its orbit due to the SN event of 1181?  Binaries become unbound if they suddenly lose more than half their mass, and binaries become unbound if the exploding star is given a large kick velocity by the SN explosion.  If the companion star is now far from the remnant, then {\it TESS} will not detect any periodic modulation.  However, this means of hiding the ex-companion star is not available for the models that need it the most.  In particular, the Hybrid CONe model (Model A, Kromers et al. 2015, see Section 7.1) has only 0.014 M$_{\odot}$ ejected at 1100 km s$^{-1}$, and such cannot give a significant kick.  To take an extreme case, if 0.1 M$_{\odot}$ is ejected as a narrow jet, then a Chandrasekhar-mass WD would only recoil at 78 km s$^{-1}$, and such a kick velocity is more than 4$\times$ smaller than the orbital velocity of any consequential companion, so the orbit cannot be unbound.  Further, the expanding shell is round and uniform, with no significant asymmetry, so any kick velocity must be small.  Further, the {\it Gaia} DR3 proper motion of the central star of Pa 30 is close to the motions of nearby stars with similar parallax, so the apparent transverse velocity is 15$\pm$23 km s$^{-1}$, which is much too small to unbind any companion.

{\bf (4)} An inventive modeler can always think up possibilities to hide the modulation on the orbital period, and such are hard to anticipate or test.  For example, I can imagine that the fierce and hot stellar wind being blown by the primary might be sufficiently large so as to enshroud the entire binary system.  With such contrived possibilities, I judge that unknown-theory-scenarios cannot now be used to make a loophole to avoid the strict {\it TESS} modulation limits.  

In all, the strict limits on periodic modulation from {\it TESS} rules out the possibility of a current consequential companion star, barring unlikely or contrived situations.

\section{SPECTRAL ENERGY DISTRIBUTION}

The spectral energy distribution (SED) of the central star can tell about the energetics and physics of the wind photosphere around the remnant.  Lykou et al. (2022) has constructed the SED from the optical red bands out to the far-infrared (near 0.6--160 microns) to reveal a prominent excess from around 20--160 microns that has a rough blackbody spectrum with a temperature near 60 K.  The interpretation is that this is thermal emission from dust particles that suffuse through the nebula.  The outgoing shell cannot sweep up enough dust, so the dust can only be from dust formation in the SN ejecta.  Such is expected because the ejecta has near 0.1 M$_{\odot}$ of dust-forming elements expanding at the relatively low velocity of 1100 km s$^{-1}$, and we know from experience with ordinary nova eruptions that such conditions will produce large quantities of dust.

The SED can be readily constructed with good sampling from the far-infrared to the ultraviolet, with an additional point in the X-ray regime.  The magnitudes and fluxes are given in Table 3.  The first column is the name of the band.  The fluxes in the second column are all as observed, although the X-ray flux has already been extinction corrected.  The third column is the satellite or program name, along with a reference keyed in the footnote.  The fourth column is the logarithm of the frequency, $\nu$, for the middle of the band in units of Hertz.  The last column is the logarithm of the extinction-corrected flux, $F_{\nu}$, in units of Janskys.  The extinction corrections are based on the $E(B-V)$ value of 0.84 mag (Oskinova et al. 2020).  These are plotted as the SED in Fig. 10.  These SED inputs were observed on dates spread out over the last decade, but the moderate variability cannot much effect the shape of the SED, because even a 20 per cent variability makes for only a small deviation in Fig. 10.  The spectrum does have prominent emission lines that will make for some small scatter about a straight line in Fig. 10, although the one most prominent line (O\,$\textsc{VI}$ at 3811~\AA) slips between the $B$ and $NUV$ bands.  In all, Fig. 10 shows only a small scatter in points about the best-fitting line.  

\begin{table*}
	\centering
	\caption{Spectral Energy Distribution for the Central Star}
	\begin{tabular}{lllll}
		\hline
		Band & Measured flux (units)  &   Source [Ref.]   &   Log[$\nu$] (Hz)   &   Log[$F_{\nu}$] (Jy) \\
		\hline
		
$W3$	&	11.15 (Vega mag)	&	{\it WISE} [1]	&	13.40	&	-2.92	\\
$W2$	&	12.14 (Vega mag)	&	{\it WISE} [1]	&	13.81	&	-2.56	\\
$W1$	&	12.28 (Vega mag)	&	{\it WISE} [1]	&	13.95	&	-2.34	\\
$K$	&	13.20 (mag)	&	2MASS [2]	&	14.13	&	-2.34	\\
$H$	&	13.63 (mag)	&	2MASS [2]	&	14.27	&	-2.24	\\
$J$	&	13.85 (mag)	&	2MASS [2]	&	14.38	&	-2.04	\\
$y$	&	3.69 (milliJansky)	&	Pan-STARRS [3]	&	14.49	&	-1.99	\\
$z$	&	3.16 (milliJansky)	&	Pan-STARRS [3]	&	14.54	&	-2.03	\\
$i$	&	3.37 (milliJansky)	&	Pan-STARRS [3]	&	14.60	&	-1.89	\\
$r$	&	2.21 (milliJansky)	&	Pan-STARRS [3]	&	14.68	&	-1.85	\\
$r'$	&	15.42 (mag)	&	APASS [4]	&	14.68	&	-1.87	\\
$V$	&	15.63 (mag)	&	APASS [4]	&	14.74	&	-1.65	\\
$g$	&	1.52 (milliJansky)	&	Pan-STARRS [3]	&	14.79	&	-1.57	\\
$g'$	&	15.97 (mag)	&	APASS [4]	&	14.79	&	-1.53	\\
$B$	&	16.34 (mag)	&	APASS [4]	&	14.83	&	-1.54	\\
$NUV$	&	18.56 (AB mag)	&	{\it GALEX} [5]	&	15.12	&	-1.20	\\
0.2--12.0 keV	&	1.1$\times$10$^{-12}$ (erg cm$^{-2}$ s$^{-1}$)	&	{\it XMM} [6]	&	17.38	&	-7.41	\\

		\hline
	\end{tabular}	
	
\begin{flushleft}	
References: [1] {\it WISE}: Wright et al. (2010), https://irsa.ipac.caltech.edu/cgi-bin/Gator/nph-scan?projshort=WISE  
[2] 2MASS: Skrutskie et al. (2006), https://irsa.ipac.caltech.edu/cgi-bin/Gator/nph-scan?projshort=2MASS   
[3] Pan-STARRS: Chambers et al. (2016), https://catalogs.mast.stsci.edu/panstarrs/ 
[4] APASS: Henden et al. (2009), https://www.aavso.org/aavso-photometric-all-sky-survey-data-release-1 , https://www.aavso.org/download-apass-data
[5] {\it Galex}: Martin et al. (2005), http://galex.stsci.edu/gr6/?page=mastform
[6] {\it XMM}: Oskinova et al. (2020)
\end{flushleft}
	
\end{table*}

\begin{figure}
	\includegraphics[width=1.0\columnwidth]{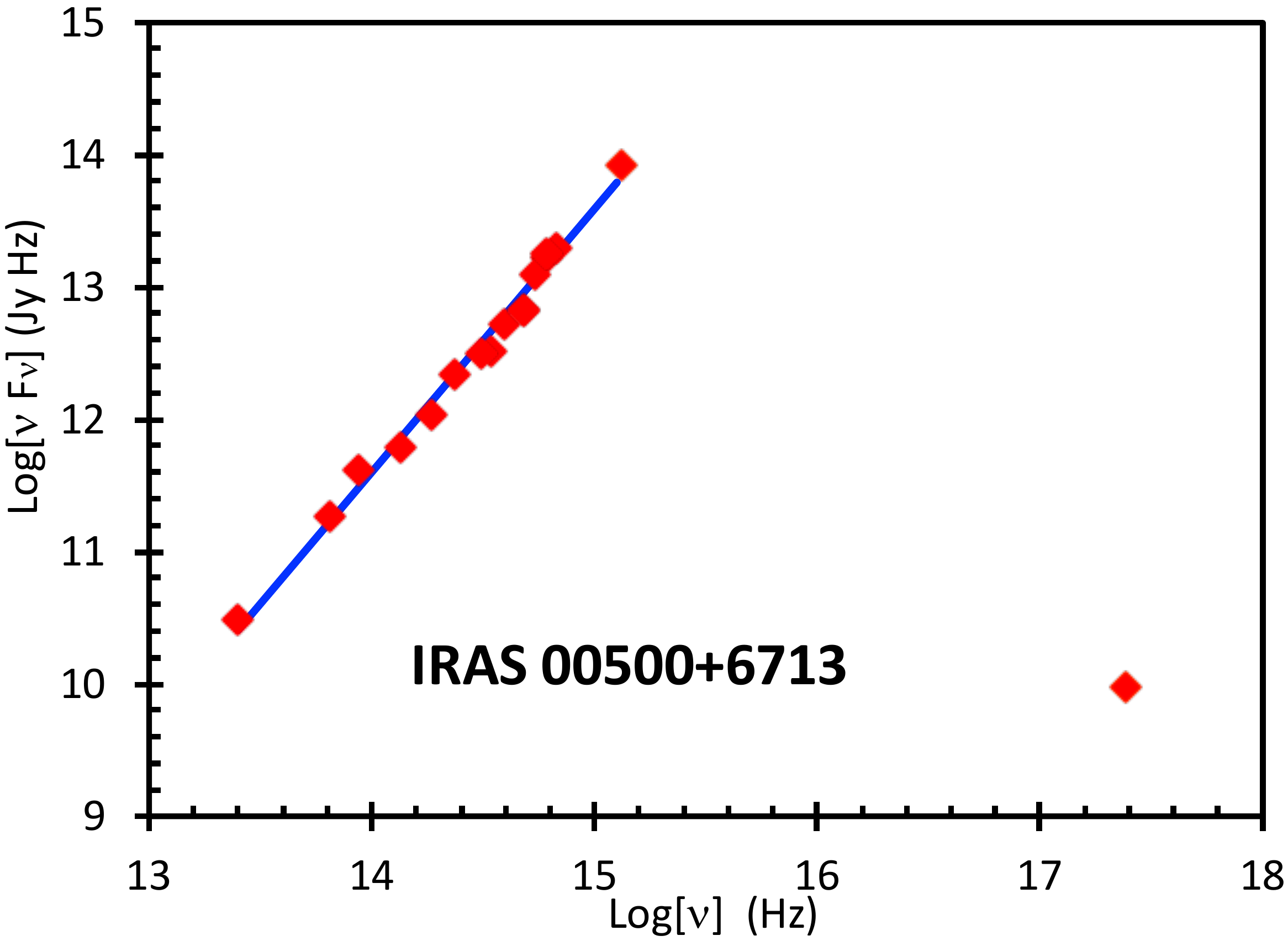}
    \caption{Spectral Energy Distribution (SED) for the central star.  This logarithmic plot of $\nu F_{\nu}$ versus $\nu$ shows where the radiant energy comes out, with most of the energy coming out in the ultraviolet and possibly into the extreme-ultraviolet.  The infrared and X-ray photons carry only a small fraction of the luminous energy put out by the remnant star.  We see that the SED is well-sampled from the far-infrared to the ultraviolet, and these data points closely follow a straight line.  That is, the remnant of SN 1181 is a good power law over a wide range of frequency, with $F_{\nu} \propto \nu ^{0.99\pm0.07}$.  The SED is far from any blackbody or any disc model.  The power-law property is a hallmark of non-thermal radiation.  The radiant energy from the far-infrared to the ultraviolet is obtained by integrating the SED under the power law, and correcting for the {\it Gaia} distance, with the luminosity equal to 128$\pm$24 L$_{\odot}$.}  
\end{figure}

The SED is closely a power-law from the ultraviolet to the far infrared.  The fitted power-law is $F_{\nu} \propto \nu ^{0.99\pm0.07}$.  The uncertainty is dominated by the uncertainty in the extinction correction, for which the Oskinova et al. (2020) value of $E(B-V)=0.84\pm0.04$ was adopted.  The small scatter points to the SED as closely and accurately being a power-law distribution, with this close to $F_{\nu} \propto \nu$.

The SED does {\it not} show the distinct shape of a blackbody, the slope is greatly different from a Rayleigh-Jeans distribution ($F_{\nu} \propto \nu ^2$), and the shape is not from any disc model (with $F_{\nu} \propto \nu ^{1/3}$ in the middle).  Good power-laws over a broad frequency range are the hallmark of non-thermal emission.  

The luminosity of the remnant can be directly measured from the SED.  The luminosity over a frequency range comes from the integral $4\pi D^2 \int F_{\nu} d\nu$.  From the far-infrared to the ultraviolet, the observed luminosity is 128$\pm$24 L$_{\odot}$.  The quoted uncertainty is dominated by not knowing the exact interstellar extinction.  

The observed remnant luminosity from far-infrared to ultraviolet is 128 L$_{\odot}$, yet there is still an unknown flux between the ends of the bands with the {\it Galex} NUV filter and the {\it XMM} band, which is from 5.5 to 200 eV.  The missing energy in the extreme-ultraviolet depends critically on how the power-law is cutoff before the low X-ray point.  If the SED cuts off sharply just above the {\it Galex} NUV point, then the missing energy is negligible.  At the other extreme, if the $F_{\nu} \propto \nu$ power-law continues closely up to the {\it XMM} X-ray band before cutting off sharply, then the missing luminosity is just over 10$^5$ L$_{\odot}$.  Both extremes are clearly unlikely.  For some sort of a middle ground, a simple power-law interpolating from the ultraviolet to the X-ray gives a missing luminosity of 107 L$_{\odot}$.  For another schematic middle ground, an extrapolation of the observed power-law out to a sharp cutoff at the Lyman limit gives a missing luminosity of 660 L$_{\odot}$.  The observed power-law must come to a strong break somewhere in the extreme-ultraviolet.  The missing luminosity is likely to be from a hundred to many-hundreds of solar-luminosities.  So the total radiant luminosity of the remnant is likely between two-hundred and many-hundreds of solar luminosities.

For just looking at the optical flux, a useful number is the absolute $V$ magnitude over recent years.  With the APASS measure of $V$=15.63, the {\it Gaia} distance of 2460 pc, and the extinction of 0.84 mag, the absolute magnitude is $+$1.07 mag.

The observed light (128 L$_{\odot}$, $M_{\rm V} = +1.07$) is certainly not dominated by any companion or ex-companion star, as the SED does not have any blackbody shape.  Further, any ex-companion cannot be a red giant or a massive star (say, a Wolf-Rayet star) because the absolute magnitude is far too faint.  (Such ex-companion stars will easily survive the nearby supernova explosion with little change in their luminosity, see Marietta et al. 2000.)  The observed light cannot be dominated by any sort of accretion disc, as shown by the SED shape.  The observed light cannot be thermal radiation from any sort of a white dwarf or neutron star because the surface area is greatly too small to provide the observed luminosity.  This leaves few possibilities for the origin of the observed light.  The only reasonable source is for the light to be coming from the hot and fierce stellar wind.  The existence of this wind is assured from the observed broad line widths (16,000 km s$^{-1}$), with nothing else able to create such sustained high velocities.  The light from this stellar wind must be coming from some non-thermal mechanism.  

%The nature of the non-thermal emission mechanism is not known.  Perhaps the $F_{\nu} \propto \nu$ power-law is an indicative clue, but I do not understand it.  Gvaramadze et al. (2019) points out the inevitability of very high magnetic fields in and around the remnant, so this opens up many possibilities involving synchrotron radiation and synchrotron self-Compton radiation.  For the observed wind velocities of 16,000 km s$^{-1}$, each oxygen nucleus is carrying energies of 21 MeV, while the entrained electrons will have kinetic energies of 0.7 keV.  With bulk velocities around 5 per cent of light-speed, we have to start worrying about relativistic effects and nuclear reactions.  The physics of the observed stellar wind is going to be hard to apply to create unique and convincing models, especially with the substantial uncertainties in the basic scenario involving the remnant{\footnote{The experience over many decades of the Gamma-Ray Burst community, in a rather similar situation, is that a vast effort by many theorists will propose many widely-divergent models, but none will be convincing.}.

\section{FROM THE CHINESE \& JAPANESE OBSERVATIONS TO A TYPE {\rm I}\lowercase {ax} SUPERNOVA}

The analysis and discussion for my five new observational results are made within the framework of the connections from the 1181 guest star all the way to a white dwarf merger of a low-luminosity Type Iax supernova.  The case that the guest star of 1181 was a supernova\footnote{Hsi (1957) included the 1181 guest star as one item a list of 11 candidate supernovae, with no explanation or discussion, while 6 of those listed are now known to not be supernovae.} was first made by Stephenson (1971).  Pa 30 was discovered in 2013 by D. Patchick (hence the "Pa" catalog designation), and the unusual nature of the central star was recognized as a very-blue central star (Kronberger et al. 2014).  Gvaramadze et al. (2019) were the first to look closely at the central star, and they recognized the extreme wind speed, composition, and temperature, while further properties were measured by Garnavich et al. (2020), Oskinova et al. (2020), Ritter et al. (2021), and Lykou et al. (2022).  Gvaramadze et al. (2019) were the first to connect Pa 30 to a merger of two white dwarfs, while Oskinova et al. (2020) first connected Pa 30 to the Type Iax supernovae and more specifically to a merger of CO and ONe white dwarfs.  Oskinova et al. (2020) first measured an age of 350--1100 years, while Ritter et al. (2021) measured an expansion age of close to 990 years. Ritter et al. (2021) first made the crucial connection from SN 1181 to Pa 30, and they recognized the importance of this connection as creating the only possible case for very detailed study of Type Iax events.

This basic set of connections, from the 1181 guest star to a specific class of supernova, has only appeared scattered through a half-dozen papers, and the chain of connections is long.  Few researchers have the expertise to access the connections from astro-history to white dwarf merger models.  So there is utility in placing the entire chain of connections in one place, accessible for workers in all fields.  For the analysis from my five new observational results (Sections 6 and 7), I can also fill in existing loopholes in the chain.  For examples, my position analysis from the Chinese and Japanese reports rejects the possibility that the SN 1181 remnant is 3C58, my light curve information can be used to reject the possibility of a nearby classical nova, and the TESS light curve provides the critical distinction between the leading models for low-luminosity Type Iax supernova.  Further, I can fill loopholes concerning the possibilities that Pa 30 is a Wolf-Rayet bubble, that more-than-one of the many Iax models are viable, and more.

In this Section, I will follow the path by addressing four questions and conclusions; `1181 AD Guest Star Was a Supernova', `Pa 30 Is a Supernova Remnant', `Pa 30 Is The Remnant of SN 1181', and `SN 1181 Was a Low-Luminosity Type {\rm I}ax Supernova'.  Table 4 provides a convenient summary of the evidences.

%\footnote{SN of the Type {\rm I}ax sub-class are 2--10 per cent of the Type {\rm I}a rate, while the low-luminosity events are $\sim$10 per cent of the {\rm I}ax sub-class, hence justifying a characterization as `uncommon' or `rare'.  With this, an attribution of SN 1181 to a low-luminosity {\rm I}ax event might be thought of as surprising.  The resolution of this surprise is that the quoted numbers for rarity are for a magnitude-limited sample where surveys look widely over the sky, whereas the SN 1181 was discovered in the volume-limited sample of our Milky Way.  Srivastav et al. (2022) report on a volume limited sample of transients, with low-luminosity {\rm I}ax supernovae constituting 12$^{+14}_{-8}$ per cent of the Type {\rm I}a rate.  So out of the five recognized historical SNe, it is no surprise that one of them is a low-luminosity {\rm I}ax event.}

\begin{table*}
	\centering
	\caption{Path from Guest Star of 1181 to a Supernova of Type {\rm I}ax}
	\begin{tabular}{lll}
		\hline
	Conclusion  &  Evaluation & Evidence \\
		\hline
1181 AD Guest Star was a SN: \\		
 & \checkmark	&	Discovered within five days in three widely separated empires  \\
 & \checkmark \checkmark \checkmark	&	Visible 185 days, not moving  \\
 & \checkmark \checkmark	&	Small area inside {\it Chuanshe} has a $\sim$842 year old supernova remnant  \\
Pa 30 is an SNR: 	\\
 & \checkmark \checkmark \checkmark	&	The only way to get $\sim$0.01 M$_{\odot}$ of neon into a 1100 km s$^{-1}$ shell is from a supernova	\\
 & \checkmark \checkmark \checkmark	&	Shell: $V_{\rm ejecta}$=1100 km s$^{-1}$, mass $\sim$0.1 M$_{\odot}$, temperature 1--20 $\times$10$^6$ K, $M_{\rm V}$=$+$1.07	\\
Pa 30 is the remnant of SN 1181:	\\	
 & \checkmark \checkmark	&	Pa 30 is the only possible SNR inside the small SN 1181 region inside {\it Chuanshe}	\\
 & \checkmark \checkmark	&	SN 1181 had 0.0$>$$V_{\rm peak}$$>$$-$1.4, Pa 30 had $+$1.5$>$$V_{\rm peak}$$>$$-$1.5	\\
 & \checkmark \checkmark	&	Three age measures for remnant: 990$^{+280}_{-220}$, 350--1100, and $\sim$1000 years	\\
SN 1181 was a low-luminosity SN {\rm I}ax:	\\		
 & \checkmark \checkmark \checkmark	&	Remnant $L$ = 128 L$_{\odot}$	\\
 & \checkmark \checkmark \checkmark	&	Shell mass $\sim$0.1 $M_{\odot}$	\\
 & \checkmark \checkmark \checkmark	&	Shell expansion velocity is 1100 km s$^{-1}$	\\
 & \checkmark \checkmark	&	$-$15.4 $<$ $M_{{\rm V,peak}}$ $<$ $-$14	\\

% \sidehead{This is a cut-in headQQQQQQQQQQQQQQQQQQQQQQQQQ} \\
% \cutinhead{This is a cut-in headQQQQQQQQQQQQQQQQQQQQQQQQQ} \\ 
		\hline
 	\end{tabular}	
\end{table*}

\subsection{1181 AD `Guest Star' Was a Supernova}

The Guest Star of 1181 was discovered on August 6 in southern China, on August 7 in Japan, and on August 11 in northern China.  {\it Three} independent discoveries within five days is remarkable.    This triple discovery proves that the phenomenon was a real astronomical event.  This triple discovery points to the transient having a fast rising light curve around the time of the first discovery.  With the position being far from the Sun, the fast rise argues against the Guest Star being a comet.  This also points to the transient getting to $V$=0, or possibly much brighter.  Amongst candidate astronomical phenomena, only novae and supernovae can appear so bright and have a fast rise time.

The Guest Star of 1181 had a total duration of visibility of 185 days, and it did not move across the sky.  This is the proof that the Guest Star was a supernova.  Out of all known transient astronomical phenomena, only a galactic supernova can produce such a long-lasting stationary light source.   `Great comets' can rarely be visible for 185 days and longer{\footnote{The only two examples I know about are both famous; the Great Comet of 1811 and Comet Hale-Bopp in 1997 (Schaefer 1997).}, but they always must move across long paths in the sky.  The explicit consistency in celestial position from 1181 August 7 to September 26 (50 days), plus the lack of stated motion or positional change all the way out to 1182 February 6, is proof against any comet or inner Solar System body.  For stellar transients, only a nova or supernova can get brighter than $V$=0, but novae do not stay naked-eye for 185 days.  In practice, bright novae have durations of visibility from a week to two-months (CS1977; SG2002; Strope, Schaefer, \& Henden 2010).  Nevertheless, if a nova system were close enough to Earth, it could be visible for 185 days or longer.  Only two known nova had a naked-eye visibility duration of longer than 120 days (Strope, Schaefer \& Henden 2010; Schaefer 2022).  The first was V603 Aql in 1918 at a distance of 324 pc, peaking at $-$0.5 mag, and visible above $V$=5.5 for 158 days.  The second was RR Pic in 1925 at a distance of 501 pc, peaking at $+$1.0 mag, and naked eye visible for 371 days.  So it is not impossible for a nova to have a naked-eye duration of visibility of 185 days or longer.  But any such nova would necessarily be close to Earth, and prominent in quiescence in several ways, including as a bright X-ray source and as one of the brightest cataclysmic variable stars.  No such record-breaking-close old-nova is seen anywhere near the constellation {\it Chuanshe}, so the 1181 Guest Star cannot be a nova.  Taken together, the 185 day duration, the lack-of-motion, and the lack of any bright old-nova in the area constitute a proof that eliminates all possible transients other than a supernova.

The small area inside {\it Chuanshe} (see Fig. 5) contains a $\sim$842 year old supernova remnant (see Section 6.2 and 6.3) at a distance of 2410$^{+200}_{-130}$ pc.  This means that we know that a supernova appeared brighter than $V$=0 inside {\it Chuanshe} during the Song dynasty.  The coincidence of the observed Song dynasty Guest Star in {\it Chuanshe} with the required Song dynasty supernova in {\it Chuanshe} from Pa 30 is too great to be acceptable unless there is a causal connection.  This is not a proof, but it is a strong argument.

\subsection{Pa 30 Is a Supernova Remnant}

Oskinova et al. (2020) report from X-ray spectroscopy that the composition of the Pa 30 central star is 61 per cent oxygen, 15 per cent carbon, 10$\pm$3 per cent neon, 6$\pm$4 per cent silicon, 4$\pm$2 per cent magnesium, and 4 per cent sulphur{\footnote{Based on optical spectroscopy of the central star, Oskinova et al. (2020) measured similar compositions of 80$\pm$10 per cent oxygen, 20$\pm$10 per cent carbon, and 1 per cent neon.}.  And their X-ray spectroscopy for the surrounding nebula has 72 per cent carbon, 13$\pm$6 per cent oxygen, 13$\pm$4 per cent neon, plus 2$\pm$1 per cent magnesium.  And there is no hydrogen or helium to strict limits for either the central star or the surrounding shell.  That is, roughly 0.01 M$_{\odot}$ of neon (and magnesium) has been somehow ejected into a fast expanding shell.  Such is possible only from a supernova.

The only sources in the universe with massive amounts of neon (and magnesium) are deep in the cores of massive stars or in the exposed cores that have turned into ONe white dwarfs.  A bulk abundance of the neon (and magnesium) in the shell is possible only for material that was at some time burned near the core of a massive star or is now exposed in an ONe WD.  So the central star of Pa 30 must derive from one of those two sources.  Critically, the expanding shell of Pa 30 contains 13 per cent neon (and 2 per cent magnesium), and so this gas can only have been ejected from either a massive star core or from an ONe white dwarf.  The only way to eject $\sim$0.01 M$_{\odot}$ from these deep gravity wells at velocities like 1100 km s$^{-1}$ is by a supernova explosion.

There are only three possible scenarios to get bulk quantities of neon (and magnesium) from its original source deep in a gravity well into a fast expanding shell:  First, a scenario involving a thermonuclear explosion of a WD containing a large amount of neon (and magnesium) would make for a fast-expanding shell with composition like observed in Pa 30.  To get the carbon required for a thermonuclear event and to get enough neon to spew in to the nebula, the exploding WD must be something like a hybrid CONe WD, and this explosive event is a SN.  This scenario has already been proposed as a model for low-luminosity Type {\rm I}ax SNe (Kromer et al. 2015), see Model A in Section 7.  Second, a catastrophic merger of a CO WD and an ONe WD can eject a shell, with this being a supernova.  This second scenario is already a model for low-luminosity Type {\rm I}ax SNe (Kashyap et al. 2018), the Model B of Table 5.  Third, a scenario involving a massive star near the end of its life exploding the neon (and magnesium) in it core, thus sending the material out into a fast expanding shell.  This is just a CC-SN.  Various versions of CC-SNe have already been proposed as models for low-luminosity Type {\rm I}ax SNe, see Models D and E in Section 7.  The point is that the only way to get $\sim$0.01 M$_{\odot}$ of Ne (and Mg) in to a fast-expanding shell is by a SN explosion.  This proves that Pa 30 is a supernova remnant.

The basic Pa 30 setup (a central star surrounded by a massive expanding shell of gas) has a variety of possible origins.  These scenarios are supernovae, novae, planetary nebulae, and Wolf-Rayet stars.  The nebula has a suite of properties that individually are consistent with an origin in a SN explosion (specifically a SN {\rm I}ax explosion), yet each property taken alone is also consistent with origins in other classes of nebula producers.  By looking at all the properties together, all the non-SN possibilities are confidently rejected.  Here, I will just consider the properties of the velocity of the ejecta in the shell ($V_{\rm ejecta}$$\approx$1100 km s$^{-1}$, Ritter et al. 2021), the mass in the shell of ejecta ($\sim$0.1 M$_{\odot}$, Oskinova et al. 2020), the shell temperature (1--20 million degrees K, Oskinova et al. 2020), and the moderate luminosity of the ejecting star ($M_{\rm V}$=$+$1.07, see Section 5).  Let us consider the four scenarios that can create shells:  

{\bf (1)} Supernovae eject a huge mass at high velocities and the shells remain X-ray hot for millennia.  However, all the common types of SNe have much higher masses and velocities than observed from Pa 30.  Further, one class of SNe, the low-luminosity Type {\rm I}ax events, have observed properties that are consistent with the Pa 30 properties.  So at least one class of SNe accords with the Pa 30 nebula properties. 

{\bf (2)} Novae eject shells with velocities typically from 500--6000 km s$^{-1}$.  But nova shells are cold (including dust formation), and only from 10$^{-6}$ to 10$^{-4}$ M$_{\odot}$.  So the Pa 30 shell formation cannot involve any type of nova eruption.   

{\bf (3)} Planetary nebula shell formation typically ejects 0.1--0.4 M$_{\odot}$, and this one property is consistent with the shell in Pa 30.  But the planetary nebula scenario is confidently rejected because $V_{\rm ejecta}$ is of order 10 km s$^{-1}$.  Further, the hot central star (with temperatures up to 200,000 K) will ionize the gas, which will coexist with relatively cool dust and ions, yet in no case will the gas be X-ray bright or 1--20 million degrees K.  Further, the shells of planetary nebulae are mostly hydrogen and helium.  

{\bf (4)} Wolf-Rayet stars are hot and luminous at the centre of fast expanding bubble nebulae that can look similar to Pa 30.  The Wolf-Rayet stars have initial masses of $>$25 M$_{\odot}$ that have evolved off the main sequence, only to have the outer shells of hydrogen and helium gas to be stripped, leaving a core with carbon and oxygen that will soon explode as a CC-SN.  Before the CC-SN, the stellar core can produce winds from 1000-2500 km s$^{-1}$ at rates of 10$^{-5}$ M$_{\odot}$ yr$^{-1}$ (Toal\'{a} et al. 2016), while the hottest stars (the `WO' stars) have abundances that make the optical spectrum dominated by O {\rm VI} lines (Crowther, De Marco, \& Barlow 1998).  Three known wind-blown bubble nebulae are hot and X-ray luminous, with values comparable to those observed for Pa 30 (Toal\'{a} et al. 2016).  This all is similar to Pa 30.  Despite these similarities, Pa 30 differs starkly from Wolf-Rayet stars (and all possible stripped massive stars) in fundamental ways:  First, it is impossible for a Wolf-Rayet star to have Ne, Mg, Si, or S near the surface in bulk quantities {\it before} any CC-SN, as observed in Pa 30.  Second, it is similarly impossible for the Wolf-Rayet star to emit a shell where the ejecta has bulk quantities of Ne and Mg, as observed for Pa 30, whereas the Wolf-Rayet bubbles are always dominated by hydrogen and helium.  Third, the core of the Wolf-Rayet stars are all necessarily blasting out energy at the rate of a supergiant star (with luminosities $>$10$^5$ L$_{\odot}$), so that the absolute $V$ magnitudes for all Wolf-Rayet stars are $<$$-$4, in strong contrast to the central star of Pa 30 having $M_{\rm V}$=$+$1.07.

Pa 30 is proven to be a supernova remnant in two ways:  First, the only way to get a bulk composition of CONe out into an ejecta shell is the catastrophic merger or explosion involving cores of massive stars or ONe WDs, and all such cases are SNe.  Second, the shell is $\sim$0.1 M$_{\odot}$ of material being ejected at near 1100 km s$^{-1}$ that is now X-ray luminous with temperatures of millions of degrees by a star now with $M_{\rm V}$=$+$1.07, and this suite of properties cannot arise from any known scenario other than a supernova.

\subsection{Pa 30 Is The Remnant of SN 1181}

Pa 30 is the only possible SNR inside the small positional region for SN 1181.  The old candidate of 3C58 has been confidently rejected on the basis of inconsistency with the Chinese positional reports.  (3C58 is also ruled out by it being certainly much too old for an event in the year 1181.)  There are no other plausible SNRs anywhere near the position for SN 1181 (see Section 2).  Indeed, this utter lack of any plausible SNRs in the area is effectively proof that no alternative exists for SN 1181{\footnote{Any $\sim$842-year-old SN of any type, other than a low-luminosity {\rm I}ax, that is near enough to get brighter than $V$=0, would necessarily produce one of the brightest X-ray and radio sources in the sky.  That no such prominent SNR is known is a strong argument that no alternative to Pa 30 exists.  Further, it proves that the SNR Pa 30 cannot be from any of the usual classes of SNe.}}.  This is a strong argument for a causal connection between SN 1181 and Pa 30.

SN 1181 had a peak magnitude of $-$1.4$<$$V_{\rm peak}$$<$0.0.  Pa 30 is a SNR that was produced by a low-luminosity Type {\rm I}ax SN.  Such supernovae have peak absolute magnitudes from $-$13.0 to $-$16.0.  For the accurately known distance and extinction for Pa 30, the peak brightness of its supernova would be from $-$1.5 to $+$1.5 mag.  That is, both SN 1181 and Pa 30 had the SN peak magnitudes inside fairly small ranges with good overlap.  This close agreement in the $V_{\rm peak}$ provides connection between SN 1181 and Pa 30.

We now have three independent measures of the age of the Pa 30 remnant:  Ritter et al. (2021) calculated an expansion age for the Pa 30 shell as 990$^{+280}_{-220}$ years.  Oskinova et al. (2020) derive an age of 350--1100 years, as based on the interaction of the fast wind blowing a bubble inside the shell.  In Section 3.5, I find a fading-age for the Pa 30 central star of $\sim$1000 years.  Taken together, the age of the Pa 30 remnant is roughly 770--1100 years old.  This puts the year of the SN to be from roughly 920 to 1250 AD, which largely coincides with the Northern and Southern Song dynasties.  Independently, the age for SN 1181 is currently 842 years, in the middle of the Southern Song dynasty.  This coincidence between the ages of Pa 30 and SN 1181 provides a direct connection from supernova to remnant.

Three improbable coincidences connect SN 1181 and Pa 30.  Both SN and SNR are from a small region in the constellation {\it Chuanshe}, both had $V_{\rm peak}$ within 1.5 mag of $V$=0.0, and both have ages $\sim$842 years.  Individually, each of the three coincidences provides a strong argument for connecting SN 1181 with Pa 30.  With all three coincidences taken together, the likelihood is unacceptably small unless there is a causal connection.

\subsection{SN 1181 Was a Low-Luminosity Type {\rm I}ax Supernova}

The SN 1181 stellar remnant Pa 30 is luminous at $L$=128 L$_{\odot}$, and this immediately rules out all of the common types of SNe, and many of the uncommon classes.  That is, CC-SN leave behind neutron stars and black holes, and such cannot appear 842 years later like the central star of Pa 30, while the thermonuclear Type {\rm I}a explosions completely blow up the WD, leaving behind no remnant.  These common models can leave behind a surviving companion star, which might or might not have supplied mass through accretion, but in all cases the ex-companions cannot look like the central star of Pa 30.  The same holds true for all uncommon classes of SNe that I am aware of -- except for the SN {\rm I}ax class.  Some of the proposed scenarios for low-luminosity {\rm I}ax systems leave behind a luminous remnant (see the next Section).  So the single fact (that Pa 30 has a 128 L$_{\odot}$ stellar remnant) immediately proves that SN 1181 was a SN {\rm I}ax event.

Oskinova et al. (2020) measured the shell mass to be $\sim$0.1 M$_{\odot}$.  Amongst supernovae, this is greatly smaller than nearly all known classes.  That is, Type {\rm I}a events eject nearly 1.4 M$_{\odot}$, while CC-SNe eject much more mass.  Indeed, this observed ejecta mass rules out all the common and uncommon SN types -- except for Type {\rm I}ax.  Only SN {\rm I}ax have such `small' ejecta masses, both observationally and with many of the models (see Section 7).  As such, the small shell mass of Pa 30 is pointing to SN {\rm I}ax.

Ritter et al. (2021) measured the highest expansion velocity of Pa 30 to be $\approx$1100 km s$^{-1}$.  For the outer shell radius of 100 arc-secs and an age of 842 years, the {\it average} velocity is 1360 km s$^{-1}$, which is close to the current observed velocity, so the initial ejection velocity, $V_{\rm ejecta}$, must be $\sim$1500 km s$^{-1}$.  This $V_{\rm ejecta}$ is in stark contrast to all the common and uncommon SNe with velocities of order 10,000 km s$^{-1}$ -- except for Type {\rm I}ax SNe.  A slow $V_{\rm ejecta}$ is part of the definition of the {\rm I}ax class.  Observed $V_{\rm ejecta}$ range from 1000--7000 km s$^{-1}$ (Jha 2017; Karambelkar et al. 2021), while models point to the low-luminosity events as having velocities as low as 1000 or 500 km s$^{-1}$ (see Section 7).  So the observed $V_{\rm ejecta}$ proves that Pa 30 cannot come from any SN class other than low-luminosity {\rm I}ax SNe.

SN 1181 had a peak absolute magnitude of  $-$15.4 $<$ $M_{{\rm V,peak}}$ $<$ $-$14.0.  This immediately eliminates any possibility of a Type {\rm I}a SN (with peaks around $-$19.3 mag), eliminates most CC-SN (with peaks more luminous than $-$16), eliminates all the super-luminous SNe, eliminates most of the rare SN classes, and eliminates most of the SN {\rm I}ax.  All that are left are 15 per cent of the CC-SN, 10 per cent of the SN {\rm I}ax, and perhaps a small collection of rare SN classes.  This does not prove that SN 1181 is in the {\rm I}ax class, but it does eliminate most of all other SN possibilities.  As a SN {\rm I}ax event (see above), it does prove that SN 1181 is in the  low-luminosity sub-subclass,

The four measured properties ($L$, ejecta mass, $V_{\rm ejecta}$, and $M_{{\rm V,peak}}$) provide proof that SN 1181 cannot be any type of SN -- except for SN {\rm I}ax.  Rather, these properties are characteristic and part-of-the-definition for SN {\rm I}ax.  The faint value of $M_{V,peak}$ is in the lowest fraction of the {\rm I}ax luminosity distribution.  This is a convincing proof that SN 1181 was a low-luminosity SN {\rm I}ax event.

\section{THE EXPLOSION MECHANISM FOR SN 1181}

Various explosion mechanisms have been advanced for the {\it low-luminosity} SNe {\rm I}ax events.  Karambelkar et al. (2021) list models including `partial deflagration of a hybrid CONe white dwarf (Kromer et al. 2015), merger of a CO and ONe white dwarf (Kashyap et al. 2018), a helium nova (McCully et al. 2014a), an ultra-stripped electron-capture SN (Pumo et al. 2009), and a fallback massive star SN (Moriya et al. 2010).'  For convenience, I have labelled these five models as Models A--E, in order.  To this list, we can add the accretion induced collapse model (Metzger et al. 2009) labelled `Model F'.

This array of six widely divergent proposals, just to explain the explosion mechanism of the low-luminosity members of a subclass of SN, teaches several lessons:  First, no one has any confident idea as to the dominant mechanism for low-luminosity SN {\rm I}ax explosions.  Second, with many of these six mechanisms likely working in the Universe, the set of observed low-luminosity SN {\rm I}ax events are possibly, or likely, of inhomogeneous origin, with two or more mechanisms operating to make most of the observed events.  Third, with this, we have a warning that conclusions made concerning the explosive mechanism of SN 1181 should not be forced on to other individual supernovae, or on to Type {\rm I}ax SNe in general.  Fourth, many widely divergent scenarios are convergent to produce similar supernovae, and this warns us that the explosions are producing a sort of `amnesia', where much of the cause and evidence of the explosion are destroyed by the explosion.  Fifth, nevertheless, differences in the ejecta and remnant are expected between models, and this is the point to address the SN 1181 evidence for deciding the explosion mechanism for this one Type {\rm I}ax SN.

We now have enough data to select amongst the six models.  Table 5 provides a summary of the many evidences along with the evaluation for each of the six models.

\begin{table*}
	\centering
	\caption{Evidence and evaluation$^a$ of six models for explosion mechanism as applied to SN 1181 and Pa 30}
	\begin{tabular}{lllllll}
		\hline
	   &  Model A & Model B  &  Model C  &  Model D  &  Model E  &  Model F \\
	   	&	CONe WD	&	CO \& ONe WDs	&	Helium nova	&	EC-SN	&	Low-L CC-SN	&	AIC	\\
%	&	Kromer et al. (2015)	&	Kashyap et al. (2018)	&	McCully et al. (2014a)	&	Pumo et al. (2009)	&	Moriya et al. (2010)	&	Metzger et al. (2009)	\\
		\hline
Supernova:  $-$15.4$<$$M_{{\rm V,peak}}$$<$$-$14	&	\checkmark\checkmark	&	\checkmark	&	XXX	&	\checkmark	&	\checkmark	&	\checkmark	\\
Supernova: 185 days above $M_{\rm V}$=$-$10	&	X?	&	\checkmark?	&	XXX	&	?	&	\checkmark	&	XXX	\\
Central star: $M_{\rm V}$=$+$0.7	&	\checkmark	&	\checkmark	&	\checkmark	&	XXX	&	XXX	&	XXX	\\
Central star: No H or He, mostly C, O, Ne, some Mg, S, Si	&	\checkmark\checkmark	&	\checkmark\checkmark	&	XXX	&	XXX	&	XXX	&	XXX	\\
Central star: Not a binary	&	XX	&	\checkmark\checkmark	&	XX	&	\checkmark	&	\checkmark	&	\checkmark	\\
Nebula: Composition C-72\%, O-13\%, Ne-13\%, Mg-2\%	&	\checkmark?	&	\checkmark\checkmark	&	XXX	&	\checkmark\checkmark	&	\checkmark\checkmark	&	XXX	\\
Nebula: $\sim$0.1 M$_{\odot}$	&	\checkmark\checkmark	&	\checkmark\checkmark	&	XXX	&	\checkmark	&	\checkmark	&	\checkmark	\\
Nebula: $V_{\rm ejecta}$$\approx$1100 km s$^{-1}$	&	\checkmark	&	\checkmark	&	\checkmark	&	?	&	\checkmark	&	XX	\\

		\hline
 	\end{tabular}	
	
\begin{flushleft}	
$^a$Evaluations of the evidence:  A single \checkmark means that the observed evidence is consistent with the model prediction.  A double check-mark (\checkmark\checkmark) means that the model makes a prediction that nicely matches the observed case..  A question-mark means that the model has not been presented with enough information to know what is being predicted with regard to the evidence on the line.  A single `X' indicates my evaluation that the model prediction is inconsistent with the observation, but at a level that might be correctable with ordinary variations on the model.  A double `XX' gives my evaluation that the model prediction is inconsistent with the observed evidence, and I see little chance that the model can be made to agree.  A triple `XXX' notes the cases where the observed property cannot be made to agree with the model by any means, which is to say that the model is refuted in its application to SN 1181.
\end{flushleft}

\end{table*}

\subsection{Model A; Explosion of Hybrid CONe White Dwarf}

Model A is a scenario where a hybrid CONe WD arrives near the Chandrasekhar mass by accretion, when an off-centre deflagration burns only in the carbon-rich material, with the explosion ejecting mass to look like a low-luminosity SN, while leaving behind a remnant WD (Kromer et al. 2015).  (This Model A has substantial problems with the convective mixing, both at the time of formation and during an interval of simmering before ignition, such that the hybrid 2-layer structure is unlikely, with Model A then not making even a weak SN explosion, see Lecoanet et al. 2016 and Schwab \& Garaud 2019.)  This is essentially a single-degenerate Type {\rm I}a supernova that has failed to burn much of the WD because much of mass is of ONe composition.  As compared to normal Type {\rm I}a explosions, a CONe WD progenitor has little carbon to burn, so the explosion has a low luminosity and fails to destroy the WD.  A hybrid CONe WD is where the inner core is like that of a CO WD, while the outer layer is like that of a ONe WD, hence the name `hybrid'.   A hybrid CONe WD is made when its progenitor star has enough mass to create a CO core that starts the burning to form a ONe layer, but does not have enough mass to have this burning going all the way to the core.  Kromer et al. (2015) model the starting case with a 0.2 M$_{\odot}$ core equally of carbon and oxygen, surrounded by a layer with 1.1 M$_{\odot}$ made up nearly equally of oxygen and neon, and all surrounded by a layer of accreted material taking the WD mass up to the Chandrasekhar mass, which they have modeled as equal parts carbon and oxygen.  There must necessarily be a companion star to provide the accreted material to get the WD to near the Chandrasekhar mass, and this companion will survive the SN event largely unscathed.

Kromer et al. (2015) only presented results for one configuration of the CONe WD, and they had to ignite a deflagration by hand from five sites around the centre.  The 0.2 M$_{\odot}$ CO core will rapidly burn, propagating as a deflagration front.  When the burning reaches the ONe layer, the burning is quenched, even while the explosion shock waves continue to the surface.  The model calculation only allows 0.014 M$_{\odot}$ to escape as ejecta, with a composition of nearly equal parts carbon and oxygen, plus substantial fractions of nickel, silicon, and neon.  The calculated light curve will depend on the viewing angle of this asymmetric explosion, with M$_{{\rm V,peak}}$ varying from $-$14.2 to $-$14.8.  No light curve is presented out to 185 days, but it is stated that the light curve evolves significantly faster than SN 2008ha.  The stellar remnant will be a WD composed mostly of carbon, oxygen, neon, plus substantial amounts of iron-group elements, including 0.02 M$_{\odot}$ of $^{56}$Ni.  The normal radioactive decay of $^{56}$Ni has a half-life of 6.1 days, but this usual mode of decay is by electron capture, whereas nickel in a completely ionized environment will have a greatly longer half-life.  Shen \& Schwab (2017) calculate that a WD with a large mass of radioactive nickel will drive a heavy and fast wind, for which after 842 years will have typical luminosities of around 100 L$_{\odot}$.  The remnant WD will be near the Chandrasekhar mass, with its donor star still in close orbit.

How does Model A compare with the observations of SN 1181 and Pa 30?  Largely, Model A does a good job of reproducing the fundamental properties:  First, the basic scenario of a failed SN results in a low-luminosity eruption that ejects only fractions of a solar mass at a relatively low velocity.  Second, Model A predicts the composition of the ejecta to be largely C, O, and Ne.  Third, the model scenario will produce a WD stellar remnant with composition largely of C, O, and Ne, while still having enough long-lived radioactive nickel to power the strong stellar wind which should now have a luminosity comparable to that observed.  With this, Model A appears to be an excellent explanation for the unique properties of Pa 30 and SN 1181.

The matches between the Model A values and the observations are not perfect.  In particular, the ejecta mass is low by an order of magnitude, while the model duration of visibility is apparently substantially low.  And the detailed predicted abundances in the ejecta have missed out on the dominance of carbon over oxygen, has missed out on the relatively high abundance of neon, and has missed in predicting large amounts of nickel in the ejecta.  Kromer et al. reported on only one particular model, with no attempt to span any reasonable parameter space so as to match the SN 1181 properties.  I expect that the model/observation differences can be substantially improved for some set of parameters, yet to be explored.  For example, the modeled ejecta mass can be easily raised to $\sim$0.1 M$_{\odot}$ by simple increasing the mass in the progenitor's CO core.  Kromer et al. make the same point that a thorough investigation of parameter space is need.  So, I judge that these moderate differences in some of the observed/predicted values are not a significant argument against Model A.

There is {\it one} significant difference between Model A and observations, that Model A requires that the remnant WD still have the donor star in relatively close orbit, whereas the observations rule out any companion star.  That is, the strict {\it TESS} upper limits on periodic modulations has effectively ruled out the possibility of a companion with period shorter than 10 days, while the observed absolute magnitude rules out all sub-giant and giant companion stars at any distance.  In Section 4.3, I have listed various attempts to hide a companion star, but all of these have been shown to be ineffective or  unacceptably unlikely.  In the end, this one point, that the central star of Pa 30 has no reasonable means to hide the required companion star, seems to provide good grounds to reject Model A for application to SN 1181.

\subsection{Model B; Merger of CO WD and ONe WD}

Model B is a scenario with an in-spiraling merger between a CO WD and a ONe WD (Kashyap et al. 2018).  This is close to a double-degenerate model for Type {\rm I}a SNe, except that the binary pair produces a relatively low energy due to not having enough carbon-rich material to detonate.  This makes for a low $V_{\rm ejecta}$, a low ejecta mass, and a sub-luminous explosion peak, which are the fundamental properties of SN 1181.

Kashyap et al. (2018) report on their simulation of a 1.2 M$_{\odot}$ ONe WD (80 per cent oxygen and 20 per cent neon) merger with a 1.1 M$_{\odot}$ CO WD (40 per cent carbon and 60 per cent oxygen).  The final in-spiral takes around 60 seconds from start to end.  The CO WD is tidally disrupted to form a hot low-density disc around the ONe WD, and mixes with the material from the ONe WD.  When the two WDs have coalesced, an off-centre hotspot within the carbon-rich disc material will initiate unstable carbon burning and carbon detonation.  With much of the high density structure having little carbon, the explosion will generate relatively small amounts of $^{56}$Ni and relatively small amounts of nuclear energy.  The merged WD will not have enough energy to un-bind the star, so a mixed-up WD, composed mostly of oxygen and neon will remain intact at the end.  An estimated 0.08 M$_{\odot}$ of material will be ejected, mostly in the polar directions, with a typical velocity of $\sim$4000 km s$^{-1}$.  The ejecta will be composed of oxygen, carbon, silicon, sulphur, and neon, in order of their masses.  The simulation results in an asymmetric ejection (but with no reason stated for the breaking of the symmetry), resulting in the surviving single WD having a kick velocity of $\sim$90 km s$^{-1}$.  On the time-scale of hours, a fraction of  the synthesized $^{56}$Ni (0.00807 M$_{\odot}$) will fall back on to the WD, with the remnant's composition dominated by oxygen, carbon, and neon in order of masses.  The light curve has $M_{{\rm V,peak}}$ equal to $-$11.3.  The fallback $^{56}$Ni will be fully ionized, and hence having a rather long decay time-scale, such that the radioactivity can keep the remnant hot with a fierce wind for many centuries (Shen \& Schwab 2017) that closely matches the observed luminosity.

Model B provides a natural explanation for all the key features of SN 1181 and Pa 30.  As a low-luminosity event, the scenario is a good setup for the faint observed $M_{{\rm V,peak}}$, the small mass of the ejecta, and the low velocity for the ejecta.  Both the central star and the nebula come from WDs, so the composition is predominantly CONe, just as for Pa 30.  There will be no binary companion.  The fallback $^{56}$Ni provides a good source of energy on the surface of the remnant WD, so as to power a hot and heavy wind that makes for the observed fading central star.  So Model B provides a good independent scenario that matches SN 1181 and Pa 30 in all the fundamental details.

Nevertheless, Model B is not perfect in some of its numerical predictions:  First, the one specific model calculated has a substantially fainter $M_{{\rm V,peak}}$ than observed.  Second, the bolometric light curve is characterized as being fast, so it is unclear whether an appropriate model can reproduce the 185 days of visibility.  Third, the predicted composition of the remnant is close to the observed abundances, but the predicted nebular abundances missed the dominance of carbon over oxygen, missed that neon makes up one-eighth of the nebular mass, and missed that silicon is much less abundant than carbon and oxygen.  For all three of these moderate quantitative differences, I expect that the differences can all be minimized when start parameters are varied.  For example, the WD masses at the time of in-spiral can be changed to add more carbon, so as to increase the luminosity and increase the duration of visibility.  I am confident that modelers can move past the one case tried by Kashyap et al., and find a case where all the differences are minimized.

The transverse velocity of the central star is 15$\pm$23 km s$^{-1}$, in contrast to the single-case calculation by Kashyap et al. (2018) for a kick velocity of $\sim$90 km s$^{-1}$.  I judge that the numerical value is a weak prediction, because the degree of asymmetry in the ejected material is a result of turbulence in a chaotic disc, with this being sensitive to fine and un-knowable details.  Similarly, for Type {\rm I}ax models, Lach et al. (2022) report that kick velocities vary from 6.9 to 369.8 km s$^{-1}$ as the fine details of the model input vary.  This study demonstrates that the kick velocity depends critically on the specific happenstance of the ignition, so the Kashyap et al. model can easily allow for low kick velocities.  Further, to test the prediction of $\sim$90 km s$^{-1}$, we must correct the observed transverse velocity by a factor $1/\sin[\theta]$, where $\theta$ is the unknown angle between the kick velocity and the line of sight.  To match the prediction, $\theta$ need only be less than 10$\degr$ for the best estimate of the kick velocity, while $\theta$ need only be less than 25$\degr$ for the one-sigma case.  So, the nominal discrepancy between the predicted kick velocity and the observed transverse velocity is not a serious objection to Model B, because the pre-SN velocity is not known with useable confidence, because the kick velocity has substantial probability of having only a small transverse component, and because I expect that many instances of Model B can have greatly smaller kicks.

In all, Model B is a good match to the observations of SN 1181 and Pa 30.  In particular, it reproduces all of the fundamental aspects of the observations.  The three moderate numerical differences are likely to be minimized when the full parameter space is explored.  So I am concluding that Model B is good.

\subsection{Models C--F}

Model C is a scenario where the low-luminosity SN Iax events are `helium novae' (McCully et al. 2014a), and not SN at all.  Only one helium nova is known, and that is V445 Pup, in which the otherwise normal classical nova event had no hydrogen in its spectrum and it suffered a deep and long lasting dust dip that is still ongoing.  V445 Pup erupted in the year 2000, peaked at $V$=8.6, had a pre-eruption brightness of $V$=14.6, and a distance of 6272 pc (with one-sigma range of 5026--9026 pc), for absolute magnitudes of $-$6.97 at peak and $-$1.0 in quiescence (Strope, Schaefer, \& Henden 2010; Schaefer 2022).  This model fails to explain SN 1181 because a helium nova cannot get up to absolute magnitudes of around $-$15, because the stellar remnant cannot have the observed composition or the ferocious wind, and because the ejected nebula cannot have the observed mass or composition.  These differences are extreme and by many orders of magnitude, so Model C is strongly rejected to explain SN 1181.

Model D is that a massive star, stripped of all its outer layers, will undergo a core collapse triggered by electron capture (Pumo et al. 2009), to explode as a so-called electron-capture supernova (EC-SN).  Model D is of a CC-SN for the lowest possible mass of the progenitor, where the initial main sequence mass must be 10.44--10.92 M$_{\odot}$ for solar metallicity.  The low mass of material burning will lead to a low-luminosity CC-SN.  It is possible for the progenitor to have lost its hydrogen and helium outer layers (much like a Type {\rm I}c SN), so the ejecta might have a CONe composition.  For the case of the progenitor with solar metallicity, stars with mass 10.44--10.46 M$_{\odot}$, all the hydrogen and helium will have been previously stripped, and the ejecta mass will be 0.01--0.2 M$_{\odot}$, with this mass and composition being a reasonable match to that for SN 1181.  This ultra-stripped EC-SN apparently can produce a supernova and an ejecta shell that might be like those seen for SN 1181 and Pa 30.  Nevertheless, this EC-SN leaves behind a neutron star remnant. This neutron star remnant can in no way look like the central star of Pa 30, so Model D is strongly rejected to explain SN 1181. 

Model E is `a fallback massive star SN (Moriya et al. 2010)'.  The idea is that an ordinary CC-SN with a relatively small explosive energy will eject only a fraction of its outer layers, with the remainder falling back, presumably into some sort of a disc.  If the progenitor star has the outer hydrogen and helium layers stripped away, then the ejecta will have a composition dominated by carbon, oxygen, and neon.  The fiducial model of Moriya et al. (2010) looks at the case where a 13 M$_{\odot}$ progenitor with solar metallicity that has been stripped down until only the CO core remains.  In this model, they initiated an explosion by inserting a kinetic piston with 1.2$\times$10$^{48}$ ergs, and then they followed the consequences.  As in the title of Moriya et al. (2010), the model is simply a CC-SN with extremely low explosion energy, where the energy was put in by-hand with no explanation.  With this simple scenario, the fiducial model calculates that 0.074 M$_{\odot}$ of mass is ejected, with this value easily being adjusted by changing the input kinetic energy.  This model has the ejecta expanding at roughly 1270 km s$^{-1}$, and the ejecta to be composed of 20 per cent carbon, 32 per cent oxygen, 14 per cent neon, 6.6 per cent magnesium, 6.3 per cent silicon, and so on.  This model would produce a nebula closely similar to that observed for Pa 30.  Further, they calculate that the light curve shape and luminosity will match that of the SN {\rm I}ax event SN 2008ha (with $M_{{\rm V,peak}}$ equal to $-$14.0), so that this model looks likely to be able to reproduce the light curve for SN 1181.  The trouble with Model E is that the CC-SN produces a neutron star as its stellar remnant, and no configuration of a neutron star can produce anything like the central star of Pa 30, so Model E is strongly rejected as the explosion mechanism for SN 1181.

Model F is `nickel-rich outflows from accretion discs formed by the accretion-induced collapse of white dwarfs' (Metzger, Piro, \& Quataert 2009).  The scenario starts with a white dwarf accreting mass (like in a single-degenerate situation) to approach the Chandrasekhar mass, spinning up to a rapid rotation prior to collapse.  When the WD starts to collapse, under some situations, like for a ONe WD, the accretion-induced collapse (AIC) will not result in a thermonuclear explosion.  For an AIC with a rapidly rotating progenitor WD, to conserve angular momentum, a disc will form just outside the proto-neutron-star surface.  In the first few seconds, the disc will be neutrino cooled and will synthesize up to 10$^{-2}$ M$_{\odot}$ of $^{56}$Ni.  This disc will start expanding at velocities of 0.1--0.2 times light speed, with a total of 0.02 M$_{\odot}$ ejected.  The presence of $\sim$10$^{15}$ G magnetic fields can produce larger ejections, up to $\sim$0.1 M$_{\odot}$.  This expanding ejecta will be heated by nickel decay and look like a low-luminosity SN.  The calculated light curve for this scenario has a $V$-band rise time of under one day, and the light curve will fall off by 7.5 mag from peak in 5 days.  The calculation is that the ejecta will predominantly be iron-peak elements, and that few intermediate-mass elements (like O, Ca, and Mg) will be present.  Metzger et al. hypothesize that if any significant amount of WD material remains far from the collapse, it could substantially slow the ejecta, provide some CONe atoms to the ejecta, and make the duration up to 10$\times$ longer.  In all cases, the light curve cannot reproduce the duration reported by the Chinese and Japanese observers, and the shell is expanding much too fast and has all the wrong composition to be like that seen in Pa 30.  Critically, Model F predicts that the stellar remnant will be a neutron star, greatly different from what is seen for the central star of Pa 30.  In all, Model F has no chance of accounting for SN 1181 and Pa 30.

\subsection{SN 1181 Explosion Mechanism}

We have SN 1181 and Pa 30 proven to come from a low-luminosity Type {\rm I}ax SN, and there are six widely divergent models.  Models C, D, E, and F are certainly not applicable to SN 1181.  So we are left with SN 1181 arising from either Model A (deflagration of a hybrid CONe WD) or Model B (merger of a CO WD and an ONe WD).  %Model A is a `single-degenerate' scenario with a companion star feeding a white dwarf by accretion until it is near the Chandrasekhar mass, while Model B is a `double-degenerate' scenario where two WDs in a close binary ultimately in-spiral to get a combined star near the Chandrasekhar mass.  

Both Model A and B do a nice job of naturally reproducing all of the primary and fundamental features of SN 1181 and Pa 30.  Both models have similar moderate differences from theory in numerical value for some properties, but such are likely to be minimized when the model inputs are optimized for the SN 1181 case.  So, in all criteria except one, both Model A and B are good explanations for SN 1181.

%(That is, both models produce low-luminosity SNe, with low-velocity and low-mass ejecta primarily of CONe composition, with a WD remnant producing a fast wind of CONe composition that even now has high luminosity.)  

The one exception is that Model A requires a close-in donor companion star while Model B has no close companion star.  This breaks the case.  With no known or effective means to hide a companion, Model A is ruled out on this one point alone.  So I conclude that SN 1181 was a low-luminosity Type {\rm I}ax SN whose explosion mechanism was the binary merger of a CO WD and an ONe WD.

\section{CONCLUSIONS}

One decade ago, the default idea was that 3C58 was the remnant of SN 1181, despite the severe problems with its age.  Fortunately, the dogged perseverance and ingenuity for the nebula search by Patchick redeemed the situation by finding Pa 30.  Two groups (that of Gvaramadze et al. 2019 and Oskinova et al. 2020), plus the group of Ritter et al. 2021) put together the basic picture connecting SN 1181, Pa 30, and SN {\rm I}ax events.  In this paper, I am adding new results, based on written histories from a millennium ago, based on a light curve over the last century, based on photometry from the last decade and based on photometry over the last year.  My new observational results are:  {\bf (1)}	The ancient observations from 1181 AD are reporting what is a fairly small area in northern Cassiopeia, and this region contains Pa 30, while excluding 3C58 and all other SNRs. {\bf (2)} The peak $V$ magnitude was between 0.0 and $-$1.4, which translates into $-$14.5$>$$M_{{\rm V,peak}}$$>$$-$16.0.  For this conversion, I have a new distance from {\it Gaia} DR3 that places the remnant at a distance of 2410$^{+200}_{-130}$ pc.  {\bf (3) }The central star of Pa 30 is fading fast over the last 133 years, with $B$ magnitudes of 14.9 mag in 1889, to 16.2 mag in 1950, to 16.34 in 2012, and to 16.58 mag in 2022 August.  {\bf (4)} Recent light curves show aperiodic intrinsic variability with a full-amplitude of 0.24 mag.  Further, the {\it TESS} light curve place severe limits on any coherent modulation with periods from 40 seconds to 10 days as having a full amplitude of under 0.0045 mag, which strongly rules out any consequential binary companion.  {\bf (5)} The spectral energy distribution is a good power law from the far-infrared to the ultraviolet, $F_{\nu} \propto \nu ^{0.99\pm0.07}$, proving a non-thermal emission.  The observed luminosity is 128$\pm$24 L$_{\odot}$, and M$_V$=$+$1.07.

Importantly, I have collected all the data and logic into a single path running from the Chinese and Japanese observers to connect SN 1181 all the way to pointing out the explosion mechanism.  In particular, I have given proofs that the Guest Star of 1181 AD was a supernova, that Pa 30 is a SNR, that Pa 30 is the remnant of SN 1181, that SN 1181 was a Type {\rm I}ax SN, that SN 1181 was a member of the rare sub-subclass as a low-luminosity SN {\rm I}ax, and that this one SN was caused by the in-spiral merger of an ONe WD with a CO WD.

Now, SN 1181 becomes the fifth SNR for which we know the age and the SN class\footnote{The other four are Kepler's SN, Tycho's SN, Crab SN, and SN 1006.  These are all famous as the most important SNRs, teaching us much of what is known about SNRs.}.  Now we have a fifth case where we can study the remnant in exquisite detail over all wavelengths, making this case one of the few reliable connections from SN to SNR.  In our lifetimes, astrophysicists will not get any better observed case for a Type {\rm I}ax event, so our community should push hard for understanding SN 1181.

\section{ACKNOWLEDGEMENTS}

I am grateful that David Pankenier (Lehigh University) gave a thorough reading of this manuscript for issues related to the Chinese and Japanese observations.  Robert Fesen (Dartmouth College) helped substantially with detailed discussions and checking of the manuscript for issues relating to supernova remnants, plus general issues.  I am thankful that Ken Shen (University of California at Berkeley) made a critical reading of the manuscript on issues of the supernova models.  I am appreciative that the team of Dagmar Neuh\"{a}user and Ralph Neuh\"{a}user (Universit\"{a}t Jena) provided reading of this manuscript for issues related to the Chinese and Japanese observations, plus checks on the astrophysics of supernovae.  Tonny Vanmunster and Josch Hambsch (Center for Backyard Astrophysics) provided excellent photometric observations, started only one night after my request.  I thank Regina Jorgenson (Director Maria Mitchell Observatory) for sending the scanned plates.

I thank the observers and archivists of the HCO plate archives, and the DASCH program (J. Grindlay PI) for their huge and excellent effort at making high-quality scans of the individual plates available on-line.  The American Association of Variable Star Observers (AAVSO) provided a variety of useful services.  This research was made possible through the use of the AAVSO Photometric All-Sky Survey (APASS), funded by the Robert Martin Ayers Sciences Fund and NSF AST-1412587.  Funding for the {\it TESS} mission is provided by NASA's Science Mission directorate.  This paper includes data collected by the {\it TESS} mission, which are publicly available from the Mikulski Archive for Space Telescopes (MAST).  This work has made use of data from the European Space Agency mission {\it Gaia} (\url{https://www.cosmos.esa.int/gaia}), processed by the {\it Gaia} Data Processing and Analysis Consortium (DPAC, \url{https://www.cosmos.esa.int/web/gaia/dpac/consortium}).  This publication makes use of data products from the Wide-field Infrared Survey Explorer ({\it WISE}, which is a joint project of the University of California, Los Angeles, and the JPL/Caltech, funded by NASA.  This publication makes use of data products from the Two Micron All Sky Survey (2MASS), which is a joint project of the University of Massachusetts and the IPAC/Caltech, funded by NASA and the NSF.  {\it Galex} is a NASA Small Explorer mission that was developed in cooperation with the Centre National d'Etudes Spatiales of France and the Korean Ministry of Science and Technology, was  launched in April 2003, and was operated for NASA by Caltech under NASA contract NAS-98034.  The Pan-STARRS1 Survey is supported by fourteen organizations in six nations, plus two funding organizations.  The Zwicky Transient Facility (ZTF) is supported by the NSF under grant No. AST-2034437 and a collaboration including Caltech, IPAC, the Weizmann Institute for Science, the Oskar Klein Center at Stockholm University, the University of Maryland, Deutsches Elektronen-Synchrotronand Humboldt University, the TANGO Consortium of Taiwan, the University of Wisconsin at Milwaukee, Trinity College Dublin, Lawrence Livermore, and IN2P3, France.

\section{DATA AVAILABILITY}

The AAVSO, APASS, DASCH, {\it Gaia}, {\it Galex}, Pan-STARRS, {\it TESS}, 2MASS, {\it WISE}, and ZTF data are publicly available on-line.  All other inputs are presented in this paper or are from the cited references.

%%%%%%%%%%%%%%%%%%%% REFERENCES %%%%%%%%%%%%%%%%%%

{}

%%%%%%%%%%%%%%%%%%%%%%%%%%%%%%%%%%%%%%%%%%%%%%%%%%
% Don't change these lines
\bsp	% typesetting comment
\label{lastpage}
\end{document}